\documentclass[twocolumn]{aastex631}

\usepackage{hyperref,enumerate,numdef}
\input{colordvi}

\hypersetup{
    unicode=false,                  
    pdftoolbar=true,                
    pdfmenubar=true,                
    pdffitwindow=true,              
    pdfstartview={FitH},            
    pdftitle={S-PLUS MP},             
    pdfauthor={vplacco},            
    pdfsubject={Astronomy},         
    pdfcreator={dvipdf},            
    pdfproducer={dvipdf},           
    pdfkeywords={metal-poor stars}, 
    pdfnewwindow=true,              
    colorlinks=true,                
    linkcolor=red,                  
    citecolor=blue,                 
    filecolor=magenta,              
    urlcolor=cyan,                  
    breaklinks=true,
    linktocpage
}

\newcommand{\abund}[2]{\ensuremath{[\mathrm{#1}/\mathrm{#2}]}}
\newcommand{\afe}{\abund{\alpha}{Fe}}
\newcommand{\cfe}{\abund{C}{Fe}}

\newcommand{\metal}{\abund{Fe}{H}}

\newcommand{\teff}{\ensuremath{T_\mathrm{eff}}}
\newcommand{\logg}{\ensuremath{\log\,g}}

\newcommand{\A}[1]{\ensuremath{A(\mathrm{#1})}}

\newcommand{\K}{\ensuremath{\mathrm{K}}}

\newcommand{\rave}{\object{SPLUS~J2104$-$0049}}

\newcommand{\splus}{S-PLUS}
\newcommand{\jplus}{J-PLUS}
\newcommand{\xplus}{J-PLUS and S-PLUS}

\newcommand{\colora}{{\texttt{(J0378-i)-(J0410-J0660)}}}
\newcommand{\colorx}{{\texttt{(J0395-J0410)-(J0660-J0861)}}}
\newcommand{\colory}{{\texttt{(J0395-J0660)-2$\times$(g-i)}}}
\newcommand{\jcaii}{{\texttt{J0395}}}
\newcommand{\jhdel}{{\texttt{J0410}}}
\newcommand{\jhalf}{{\texttt{J0660}}}

\shortauthors{Placco et al.}

\begin{document}

\title{Mining S-PLUS for Metal-Poor Stars in the Milky Way}

\author[0000-0003-4479-1265]{Vinicius M.\ Placco}
\affiliation{NSF’s NOIRLab, 950 N. Cherry Ave., Tucson, AZ 85719, USA}

\author[0000-0002-8048-8717]{Felipe Almeida-Fernandes}
\affiliation{NSF’s NOIRLab, 950 N. Cherry Ave., Tucson, AZ 85719, USA}
\affiliation{Departamento de Astronomia, Instituto de Astronomia, Geof\'isica e
Ci\^encias Atmosf\'ericas da USP, Cidade \\ Universit\'aria, 05508-900, S\~ao
Paulo, SP, Brazil}

\author[0000-0002-0544-2217]{Anke Arentsen}
\affiliation{Universit\'e de Strasbourg, CNRS, Observatoire Astronomique de
Strasbourg, UMR 7550, F-67000 Strasbourg, France}

\author[0000-0001-5297-4518]{Young Sun Lee}
\affiliation{Department of Astronomy and Space Science, Chungnam National
University, Daejeon 34134, South Korea}

\author[0000-0002-4064-7234]{William Schoenell}
\affiliation{GMTO Corporation 465 N. Halstead Street, Suite 250 Pasadena, CA 91107, USA}

\author{Tiago Ribeiro}
\affiliation{Rubin Observatory Project Office, 950 N. Cherry Ave., Tucson, AZ 85719, USA }

\author[0000-0002-2484-7551]{Antonio Kanaan}
\affiliation{Departamento de F\'isica, Universidade Federal de Santa Catarina, Florian\'opolis, SC 88040-900, Brazil}

\correspondingauthor{Vinicius M.\ Placco}
\email{vinicius.placco@noirlab.edu}

\begin{abstract}
This work presents the medium-resolution ($R \sim 1,500$) spectroscopic
follow-up of 522 low-metallicity star candidates from the Southern Photometric
Local Universe Survey (\splus). The objects were selected from narrow-band
photometry, taking advantage of the metallicity-sensitive \splus\ colors. The
follow-up observations were conducted with the Blanco and Gemini South
telescopes, using the COSMOS and GMOS spectrographs, respectively.  The stellar
atmospheric parameters (\teff, \logg, and \metal), as well as carbon and
$\alpha$-element abundances, were calculated for the program stars in order to
assess the efficacy of the color selection. Results show that $92^{+2}_{-3}\%$
of the observed stars have \metal$\leq -1.0$, $83^{+3}_{-3}\%$ have \metal$\leq
-2.0$, and  $15^{+3}_{-3}\%$ have \metal$\leq -3.0$, including two ultra
metal-poor stars (\metal$\leq -4.0$). 
The 80th percentile for the metallicity cumulative distribution function of the
observed sample is \metal$= -2.04$.
The sample also includes 68 Carbon-Enhanced Metal-Poor (CEMP) stars.
Based on the calculated metallicities, further \splus\, color cuts are proposed,
which can increase the fractions of stars with \metal$\leq -1.0$ and $\leq -2.0$
to 98\% and 88\%, respectively.
Such high success rates enable targeted high-resolution spectroscopic follow-up
efforts, as well as provide selection criteria for fiber-fed multiplex
spectroscopic surveys.
\end{abstract}


\section{Introduction}
\label{intro}

There is a wealth of information contained in the colors of stars
\citep{allende2016}, which are simply the difference in integrated fluxes on two
given photometric bandpasses. The first determinations of effective temperatures
from photometry date back to the early 20$^{th}$ century \citep{greaves1929} and
since then extensive work has been conducted to characterize and calibrate
temperature scales in optical \citep{bessell1979} and near-infrared
\citep{alonso1996,alonso1999,casagrande2010} systems, just to mention a few.
The same applies to estimating the metallicity ([Fe/H]\footnote{\abund{A}{B} =
$\log(N_A/{}N_B)_{\star} - \log(N_A/{}N_B) _{\odot}$, where $N$ is the number
density of atoms of a given element in the star ($\star$) and the Sun ($\odot$),
respectively.}) of stellar sources from photometry. Many studies in the
literature relied on the calibration of the ultra-violet excess for stellar
sources, which is heavily dependent on the metallicity
\citep{wallerstein1964,schuster1989,bonifacio2000}, but there are others that
take advantage of infrared colors, depending on the stellar population
\citep[e.g. cold brown dwarfs -][]{leggett2010}.

More recently, large-scale surveys have taken these photometric
parameter determination strategies to the next level by building databases with
millions of spectroscopically-observed objects. Two such examples are the Sloan Digital Sky
Survey \citep[SDSS;][]{york2000} in the northern hemisphere and the SkyMapper
Sky Survey \citep[SMSS;][]{wolf2018} in the southern hemisphere. Both of these
surveys conducted separate sub-surveys aiming to perform medium-resolution ($R
\sim 1,500$) spectroscopic follow-up of stars in the Milky Way Galaxy: the Sloan
Extension for Galactic Understanding and Exploration
\citep[SEGUE-1;][]{yanny2009} and SEGUE-2 \citep{rockosi2022}, and the AAOmega
Evolution of Galactic Structure \citep[AEGIS;][among
others]{keller2007,yoon2018}. These not only served as the basis for a number
of statistical studies of stellar populations in the Milky Way
\citep{ivezic2012}, but also as prime datasets for high-resolution
spectroscopic follow-up and calibration of photometric parameter
determinations.
%
%


The SDSS makes use of the Sloan Photometric System, which comprises of
five colors ($u^{\prime}/g^{\prime}/r^{\prime}/i^{\prime}/z^{\prime}$) that cover
the region between 3,000\,{\AA} and 11,000\,{\AA} into five essentially
nonoverlapping passbands \citep{fukugita1996}.
\citet{ivezic2008} were able to determine temperatures (with typical
uncertainties of $ \sim 100$~K) and metallicities (with uncertainties of 0.2~dex
or better for $-2.0 \leq $\metal$ \leq -0.5$) for over 2 million F/G stars in
the Milky Way. One of the limitations on the low-metallicity end is due to the
broadness of the $u$ filter, which loses its metallicity sensitivity, hampering
efforts to extend the determinations to \metal$\leq -2.5$.
In two follow-up studies from the work of \citet{ivezic2008}, \citet{an2013} and
\citet{an2015} re-determined the photometric metallicities and the metallicity
distribution functions (MDFs) in the Galactic halo from SDSS photometry. These efforts
relied on improved photometry from the Stripe 82 region of SDSS and were able to
increase the metallicity range of the photometric estimates down to \metal$ \sim
-2.5$. 


The SkyMapper filter set design was optimized for stellar
astrophysics, in particular the study of stellar populations in the Milky Way
\citep{keller2007}.  It is composed of six filters: $u/v/g/r/i/z$
\citep{bessell2011}. The $u$ ($\lambda_{\rm cen}=349$ nm) and $v$ ($\lambda_{\rm
cen}=384$ nm) filters are a two-filter version of the SDSS
$u^{\prime}$, providing additional photometric sensitivity.
The SkyMapper Data Release 1 \citep[DR1;][]{wolf2018} has been extensively used
to determine photometric stellar atmospheric parameters and select
low-metallicity stars for spectroscopic follow-up.
\citet{casagrande2019}, using the SkyMapper DR1, were able to determine
temperatures and metallicities with uncertainties better than $\sim 100$~K and
$\sim 0.2$~dex for \metal$\geq -2.0$, respectively.
A similar study by \citet{huang2019}, limited to red giant stars, were able to
reach slightly lower uncertainties ($\sim 80$~K and $\sim 0.18$~dex), however
with the parameter space still limited to \metal$\geq -2.0$, with only a few
objects with metallicities below this threshold. \citet{chiti2021a} extended the
low-metallicity limit to \metal$< -2.5$ with $\sigma \sim 0.31$~dex, with the
goal of constructing the photometric MDF of
the Milky Way \citep{chiti2021b}. In terms of spectroscopic follow-up,
\citet{dacosta2019} presents 2,618 candidates selected to have photometric
\metal$<-2.0$ from their metallicity-sensitive diagram. Results show that over
40\% of the observed stars have \metal$\leq -2.75$.




The Pristine Survey \citep{starkenburg2017} has been successfully using
narrow-band photometry on the metallicity sensitive \ion{Ca}{2} H and K
absorption features (in addition to SDSS broad-band $g$ and $i$) to search for
low-metallicity stars in the Galaxy from the Northern hemisphere. The $\sim
100$\,{\AA} wide narrow-band filter has a larger predictive power than the
broader band counterparts of SDSS and SkyMapper and is able to successfully
predict metallicities in the \metal$\sim -3.0$ regime \citep{youakim2017}. The
results of a three-year medium-resolution spectroscopic follow-up campaign show
that $\sim 70\%$ of the 1,007 stars observed have \metal$< -2.0$ and $\sim 9\%$
have \metal$< -3.0$ \citep{aguado2019}\footnote{The Pristine Survey has
individual photometric metallicities to compare to the spectroscopic
determinations. For the work of \citealt{aguado2019}, 23\% of the stars with
photometric \metal$< -3.0$ also have spectroscopic \metal$< -3.0$.}. 


The Javalambre Photometric Local Universe Survey \citep[\jplus;][]{cenarro2019} and
the Southern Photometric Local Universe Survey
\citep[\splus;][]{mendesdeoliveira2019} have a unique 12 broad- and narrow-band
filter set, consisting of four SDSS ($g$, $r$, $i$, $z$), one modified SDSS $u$,
and seven narrow-band filters. The narrow-band filters were designed to probe
very specific regions in the optical wavelength regime and accommodate a wide
variety of science cases, from high-precision photometric redshifts
\citep{molino2019} to the identification of low-metallicity stars in the
Galactic halo \citep{galarza2022}. The names and key absorption features sampled
by the narrow-band filters are: 
$J0378$ -- [\ion{O}{2}]; $J0395$ -- \ion{Ca}{2} H$+$K; $J0410$ -- H$\delta$;
$J0430$ -- $G$~band; $J0515$ -- Mg~$b$ triplet; $J0660$ -- H$\alpha$; and
$J0861$ -- Ca triplet.
It is worth pointing out that the $J0395$ filter shares a similar central
wavelength and width as the Pristine narrow-band filter.  However, \xplus\ have
the advantage of also performing narrow band photometry in the Mg~$b$ triplet
($J0515$) and Ca ($J0861$) triplet regions, which are also useful for
metallicity and surface gravity determinations \citep{majewski2000}.
\citet{whitten2019} used \jplus\ photometry to predict \teff\ and \metal\ using
artificial neural networks and reached uncertainties of $\sim 91$~K and $\sim
0.25$~dex for stars in the $-3.0\lesssim$\metal$\leq -0.5$. In a follow-up study
using \splus, \citet{whitten2021} were able to estimate the first photometric
carbon abundances for a sample of over 50,000 stars, with uncertanties better
than $\sim 0.35$~dex.
Finally, \citet{galarza2022} used \jplus\ photometry to predict stellar
atmospheric parameters from machine learning techniques, reaching a success rate
of 64\% in identifying stars with \metal$< -2.5$, confirmed by medium-resolution
spectroscopic follow-up.

The possibility of accurately determining stellar atmospheric parameters and
chemical abundances for large datasets drawn from photometry, and especially for
a wide range of metallicities, is fundamental in the context of studying
low-metallicity stars. Very Metal-Poor \citep[VMP - \metal$<
-2.0$;][]{beers2005} stars are the ``local'' observational probes that allow
astronomers to address questions at cosmological scales \citep{bromm2004}.
The research that was once limited to individual stars
\citep{carney1981} has been expanded to much larger samples, allowing the
investigation of relations such as the carbon-enhancement observed in metal-poor
stars \citep{lucatello2006,aoki2007}, the possible origins of the subclasses
within this group \citep{masseron2010}, their role in the chemical evolution of
the early universe
\citep{woosley1995,heger2002,meynet2006,norris2013,frebel2015}, and the
connection between the Galactic halo and low-mass dwarf galaxies accreted
within the context of hierarchical assembly \citep[][among
others]{yuan2020,limberg2021b,shank2022}.
Growing statistics on VMP stars narrow error bars and broadens our
understanding of the early stages of the chemical evolution of the Universe.

This article reports on the medium-resolution ($R \sim 1,500$) spectroscopic
follow-up of low-metallicity star candidates selected from the \splus\ Data
Release 3 (DR3). The main goal is to assess whether the metallicity-sensitive
\splus\ colors are effective in selecting metal-poor stars for spectroscopic
follow-up. Section~\ref{secobs} describes the medium-resolution spectroscopic
observations, followed by the estimates of the stellar atmospheric parameters
and abundances in Section~\ref{secatm}.  
In Section~\ref{analysis} we analyze the sensitivity of the narrow-band
photometry to the stellar parameters, the effectiveness of the \splus\ color
selection for low-metallicity stars, the distribution of carbon and
$\alpha$-element abundances, and further improvements in the color selection.
Our conclusions and prospects for future work are provided in
Section~\ref{final}.

\begin{figure*}[!ht]
\epsscale{1.15}
\plotone{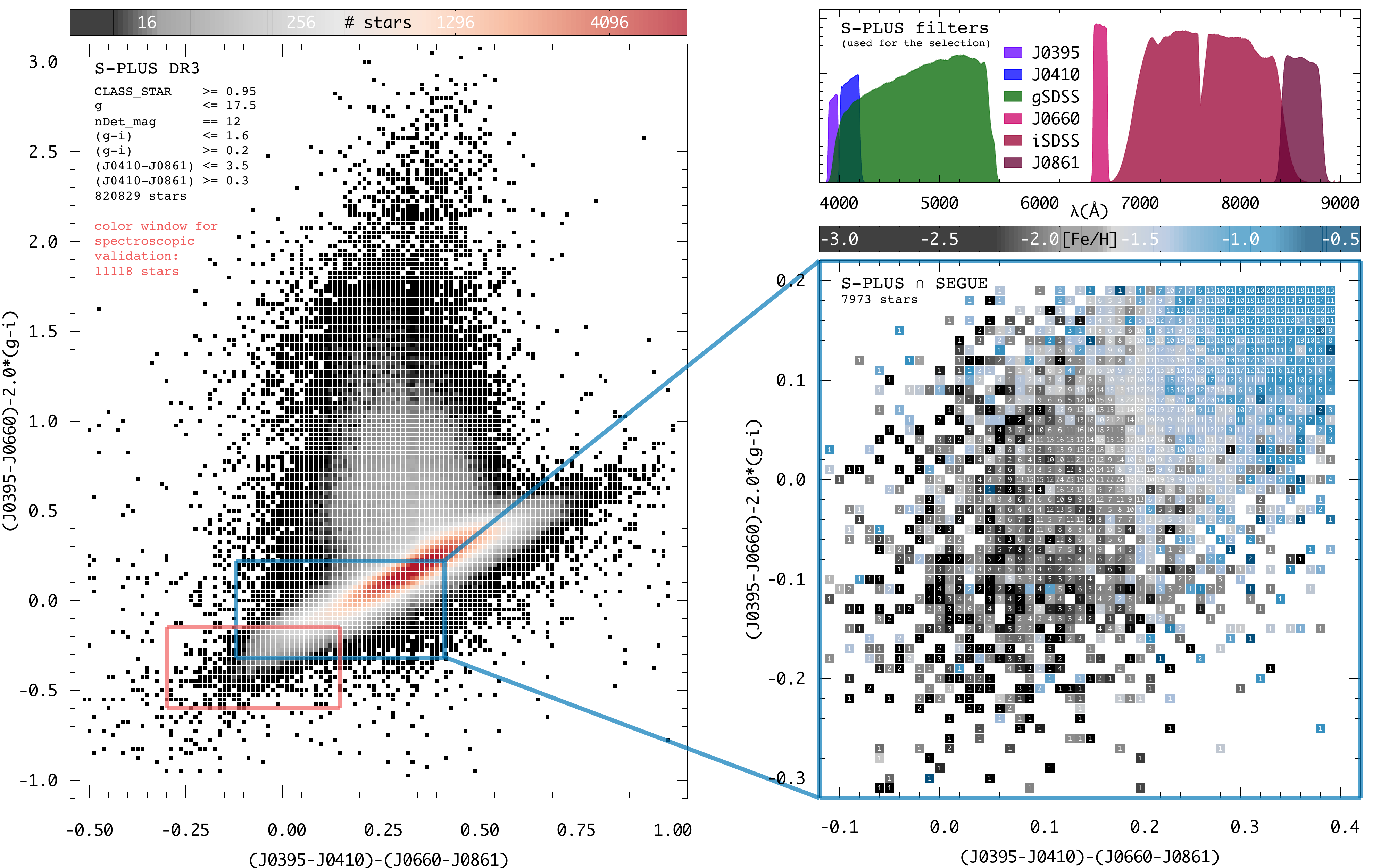}
\caption{Left panel: stellar density for the selected \protect\splus\ DR3 sample
in a color-color diagram. The red box outlines the color window for the
spectroscopic follow-up (see text for details). The inset (bottom right panel)
shows the cross-match with the SDSS/SEGUE spectroscopic database, color-coded by
the average metallicity in each bin. The number of stars in each bin is also
shown. Top right panel: \splus\ transmission curves for the six narrow-band and
broad band filters used in this selection.}
\label{photoselect}
\end{figure*}

\section{Target Selection and Observations}
\label{secobs}

\subsection{The \splus\ Data Release 3}

For this work, the \splus\ Data Release 3 (DR3; Buzzo et al., in prep) was used.
The data structure in this data release, as well as the photometric extraction
and calibration process, are the same as the S-PLUS Data Release 2
\citep[DR2;][]{almeida-fernandes2022}. The only difference is the addition of
observations in the South Galactic Hemisphere for DR3.  At the time the
candidates for this work were selected, the catalogs were only available
internally to the collaboration and are now publicly available through the
{\texttt{S-PLUS
Cloud}}\footnote{\href{https://splus.cloud/}{https://splus.cloud/}} service.

The first step in the data selection was to apply a series of restrictions in
the DR3 database, which originally contained 13,416,120 sources, mostly
related to the quality of the photometry, probability of being a stellar source,
and a color range suitable to study low-metallicity stars. Then, metal-poor star
candidates were chosen for the medium-resolution spectroscopic campaign,  based
on their location on a metallicity-dependent color-color diagram, as described
below.
The following restrictions were applied to the DR3 database:

\begin{itemize}
\itemsep-0.25em 
\item {\texttt{CLASS\_STAR}}\,$\geq 0.95$: sources having a high probability of
being a star;
\item {\texttt{gSDSS\,$\leq 17.5$}}: brightness limit for spectroscopic
follow-up within reasonable exposure times;
\item {\texttt{nDet\_magPStotal\,$= 12$}}: only sources with all twelve magnitudes
measured;
\item {\texttt{(gSDSS-iSDSS)\,$\in$\,[0.2:1.6]}}: color window to remove
possible contamination from white dwarfs and A-type stars on the blue end and
objects cooler than \teff$\sim 4000$\,K on the red end \citep[see Figure 3
in][]{yanny2009};
\item {\texttt{(J0410-J0861)\,$\in$\,[0.3:3.5]}}: same as above using a
narrow-band color;
\item {\texttt{Total}}: 820,829 stars.
\end{itemize}

The \splus\ magnitudes used throughout this work are the 3-arcsec aperture
corrected values, labelled {\texttt{PStotal}}. The left panel of
Figure~\ref{photoselect} shows the density of the selected stars in a
color-color diagram. The color \colory\ was chosen based on the work of
\citet{starkenburg2017} for the Pristine Survey, which is proven to have a
strong \metal\ dependency. The $g$ filter was replaced with \jhalf\ as a
temperature sensitive feature and the {\texttt{-2$\times$(g-i)}} was used to
re-shape the color-color diagram. In the x-axis, both \citet{starkenburg2017}
and \citet{dacosta2019} employ the {\texttt{(g$-$i)$_0$}} color. Due to the
dependency of the \ion{Ca}{2}~K line strength with temperature, there is a color
range, roughly {\texttt{(g$-$i)$_0$}$\lesssim 0.3$}, where the
metallicity-dependent color difference between a star with \metal=$-2.0$ and
$-4.0$ becomes smaller than the typical uncertainties in the photometry. In an
attempt to address this degeneracy, this work employs the \colorx\ combination,
that uses the filters centered on H$\delta$ and H$\alpha$ and provides better
temperature sensitivity. This temperature-dependent metallicity color index
should increase the success rate of finding metal-poor stars. The transmission
curves for the six \splus\ filters used in this selection are shown in the top
right panel of Figure~\ref{photoselect}.

The catalog generated from the \splus\ DR3 selection above was cross-machted with the
SDSS/SEGUE spectroscopic database. From that cross-match, sources with 
{\texttt{CLASS==QSO}}, {\texttt{$\sigma_{\rm Teff} > 200$\,K}}, and
{\texttt{S/N$ < 20$}} were excluded. The lower right panel of
Figure~\ref{photoselect} shows a section of the color-color diagram, with each $0.01
\times 0.01$ bin colored by its average \metal\ value from the spectroscopic
data. Also shown in each bin are the number of stars used to calculate the
average. The metallicity dependency is very evident in both axes and allows for
an improved selection of potential metal-poor stars for spectroscopic follow-up.

\begin{figure*}[!ht]
\epsscale{1.15}
\plotone{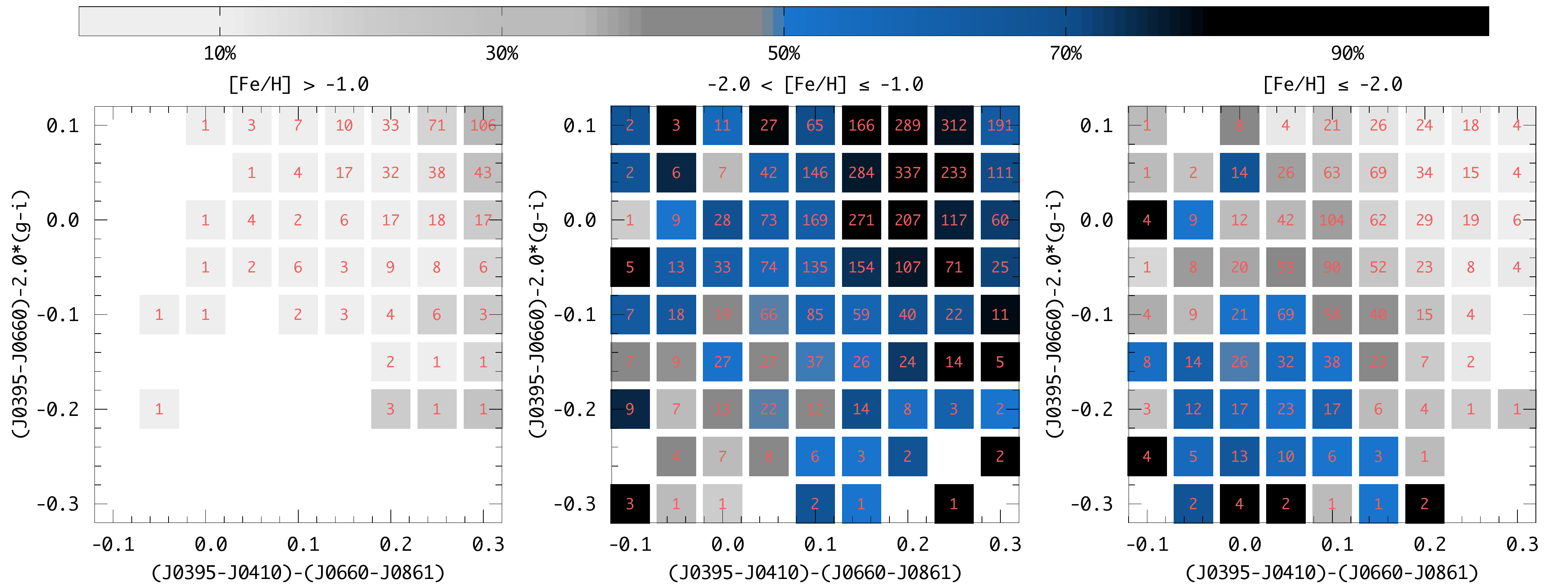}
\caption{Color-color diagram for three different \metal\ regimes. The bins in
each panel are color-coded by the fraction of stars with an average metallicity
in a given range. The number of stars in each bin is also shown.}
\label{binfull}
\end{figure*}

\begin{figure*}[!ht]
\epsscale{1.16}
\plotone{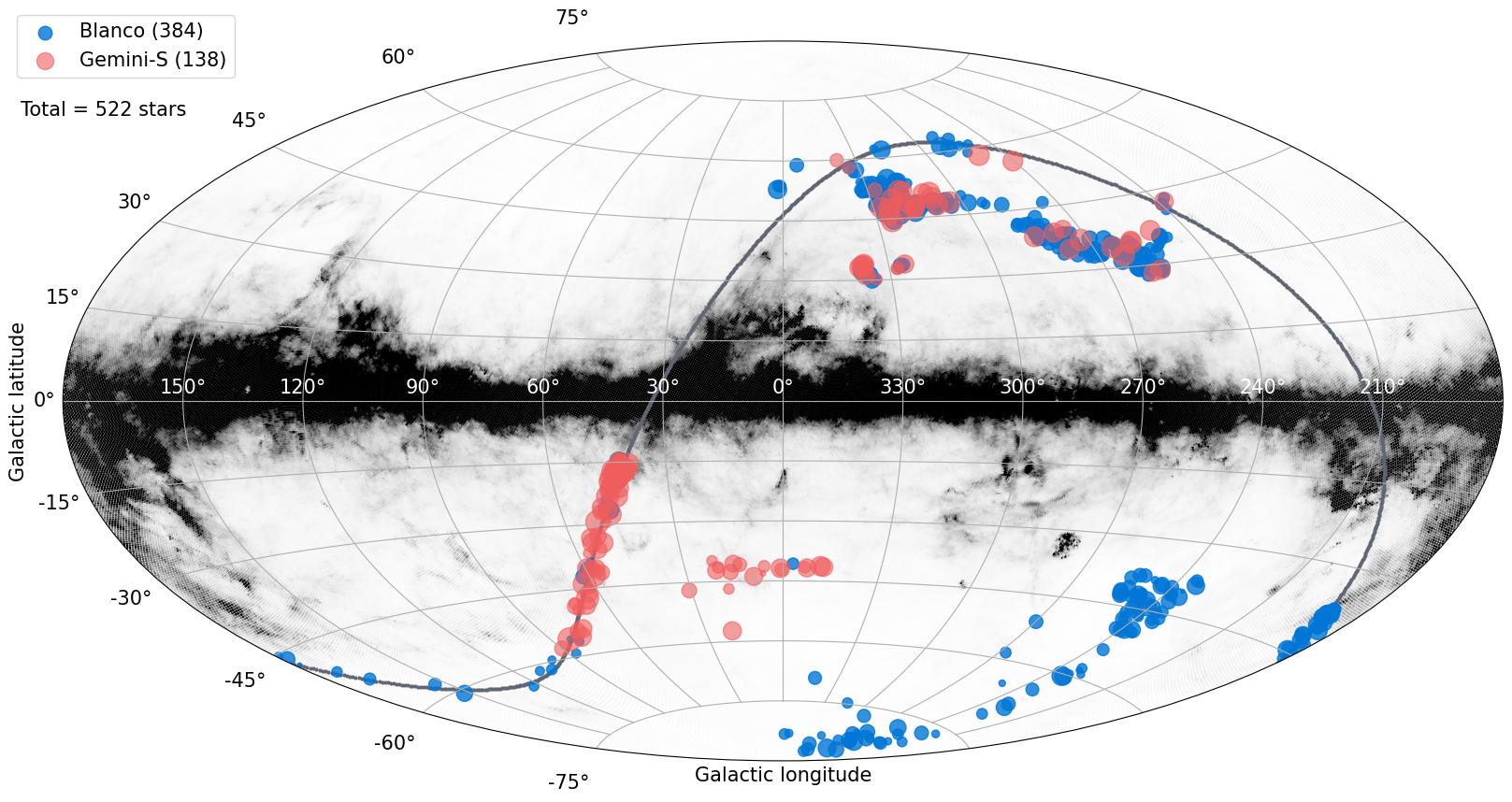}
\plotone{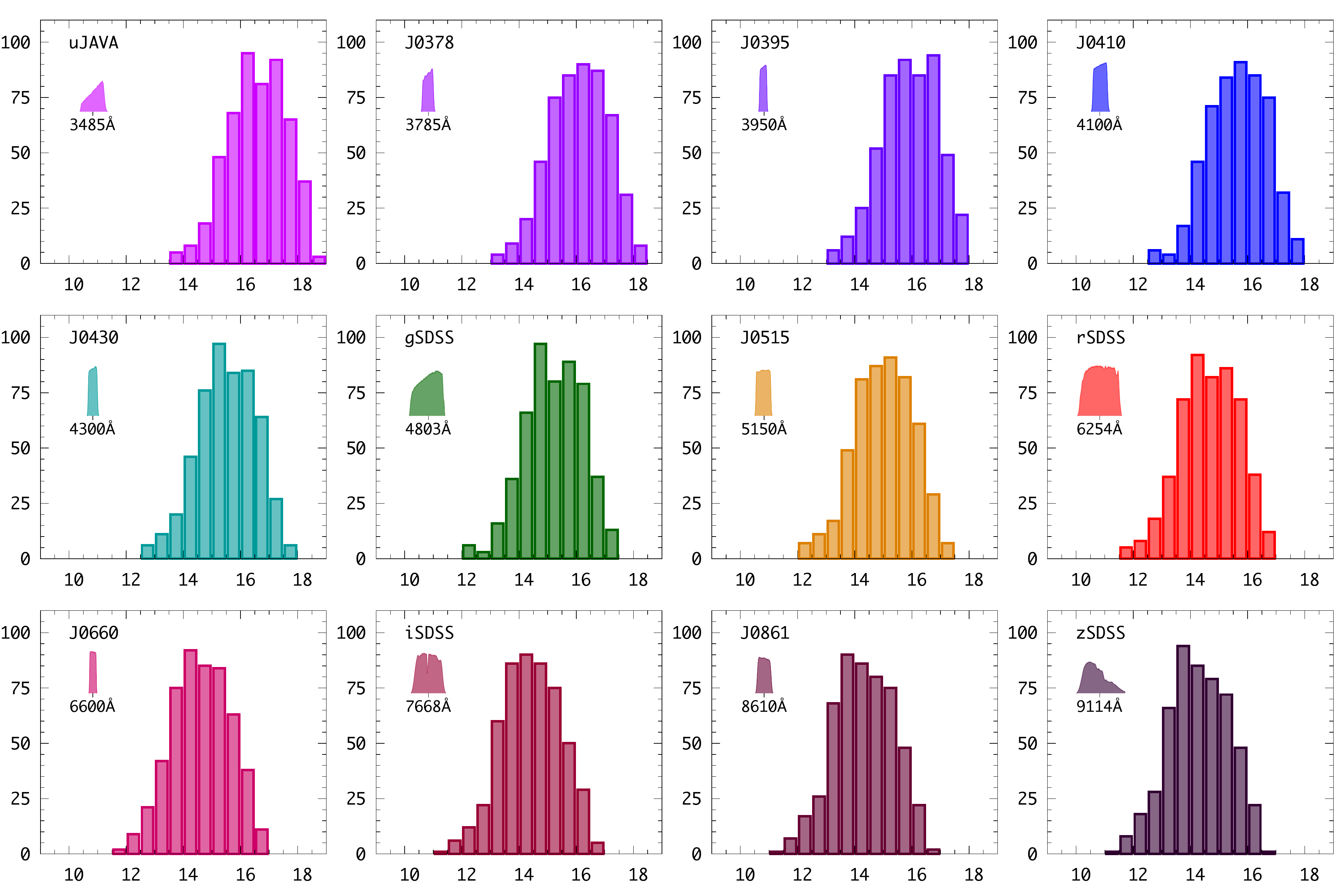}
\caption{Top: Galactic coordinates for the stars observed in this work,
color-coded by telescope. The point size is proportional to the $g$ magnitude.
The gray line traces the celestial equator. The dust map uses reddening values
from \citet{schlegel1998}. Bottom: Magnitude distribution of the observed stars
in the 12-filter system of \protect\splus. Each panel also shows the name,
transmission curve, and central wavelength (in \,{\AA}) for the filters.}
\label{footprint}
\end{figure*}

Figure~\ref{binfull} further explores this color space. Each panel shows a
different metallicity regime, color coded by the fraction of stars in each $0.05
\times 0.05$ bin. The number of stars in each bin is also shown. As an example,
the bin centered on {\texttt{(0.0,$-$0.1)}} has 1 star with \metal$>-1.0$
($\sim2\%$), 19 stars with $-2.0<$\metal$\leq-1.0$ ($\sim46\%$), and 21 stars
with \metal$\leq-2.0$ ($\sim52\%$). For the spectroscopic follow-up, from the
right panel, a cut was made where most bins have at least a 50\% fraction of
\metal$\leq -2.0$ star. Limits were also placed on the blue end of each color,
to avoid sources with potentially spurious colors. The final color window for
the selection of targets for the spectroscopic follow-up was defined as:
\colorx\,$\in$\,{\texttt{[$-$0.30:0.15]}} and
\colory\,$\in$\,{\texttt{[$-$0.60:$-$0.15]}} (see red-colored box in the left
panel of Figure~\ref{photoselect}). Within this window, which contains
11,118 stars from \splus\ DR3, targets were chosen based on their brightness and
observability at a given telescope, as well as giving preference to objects
located in the lower-left part of the selection window. Further details are
provided in the next section.

\subsection{Medium-Resolution Spectroscopy}
\label{subobs}

The spectroscopic follow-up campaign was conducted in semesters 2019A, 2019B,
2020A, 2021A, and 2022A. Data were collected for 522 metal-poor star candidates,
selected from their \splus\ photometry, described above.  The stars were
observed with two different telescope/instrument setups: Blanco/COSMOS and
Gemini South/GMOS-S.

Prior to the start of the observing campaigns in late 2018, the \splus\
candidate list was cross-matched with the {\texttt{SIMBAD}} Astronomical
Database\footnote{\href{http://simbad.cds.unistra.fr/simbad/}{http://simbad.cds.unistra.fr/simbad/}}
and stars with previously determined stellar parameters were excluded. After the
follow-up observations were concluded, another cross-match was conducted for the
522 observed targets, and four stars were found to have recent stellar parameter
determinations:
2MASSJ03145801$-$3236489 and 2MASSJ04441395$-$3356317 \citep{steinmetz2020}, 
2MASSJ11120172$-$2212075 \citep{cordoni2021}, and 
2MASSJ13103235$-$1257092 \citep{placco2019}. 
The published parameters all agree within 2$\sigma$ with the values determined in
this work (see Section~\ref{secatm} for details).

The distribution (in Galactic coordinates) of the observed stars is shown in the
top panel of Figure~\ref{footprint}, color-coded by telescope. The point size is
proportional to the $g$ magnitude of each target. The dust map traces the
Galactic plane and was constructed from the \citet{schlegel1998} reddening
values. Also shown (gray solid line) is the celestial equator. The apparent
grouping of the stars is due to the \splus\, observing strategy, which started
with the {\texttt{STRIPE82}} (equatorial) region (DR1), then moving towards halo
fields at lower southern declinations (DR2 and DR3). Note that most of the faint targets were
observed with the Gemini telescope, due to its larger aperture.
Table~\ref{telescope} lists the name, coordinates, and observing details for
each star.
The bottom panels of Figure~\ref{footprint} show the magnitude distribution of
the observed targets in all the 12 \splus\ filters. Each panel displays the
transmission curve for the filters with their central wavelength (in \,{\AA}).
Table~\ref{tsplusphoto} lists all the magnitude values (and errors) for the
observed stars, taken from the \splus\ DR3 catalog.

For consistency in the spectroscopic observations, grating/slit combinations
were chosen to yield a resolving power $R\sim 1,200-1,800$, and exposure times
were set to reach a signal-to-noise ratio of at least S/N$\sim 30$ per pixel at
the \ion{Ca}{2}~K line (3933.3\,{\AA}). Calibration frames included arc-lamp
exposures, bias frames, and quartz flats. Specific details of each instrument
and data reduction are given below.

\paragraph{CTIO Blanco Telescope}

A total of 384 stars were observed with the V\'ictor M. Blanco 4-meter
Telescope, located at the Cerro Tololo Inter American Observatory, using the
COSMOS \citep[Cerro Tololo Ohio State Multi-Object Spectrograph;][]{martini2014}
instrument.
Observations were conducted in remote visitor mode in October 2019, December
2020, January 2021, and April 2022 (2019B-0069, 2020A-0032, and 2022A-210002).
The exposure times ranged from 90 to 1800 seconds, with a total of 57.96 hours
on target.
The setup included a 600~l~mm$^{\rm{-1}}$ grating (blue setting) and a
1$\farcs$5 slit, resulting in a wavelength coverage in the range
[3600:6300]\,{\AA} at resolving power $R \sim 1,800$. 
All tasks related to spectral reduction, extraction, and wavelength calibration
were performed using standard
IRAF\footnote{\href{https://iraf-community.github.io/}{https://iraf-community.github.io/}.}
packages.

\paragraph{Gemini South Telescope}

138 stars were observed with the 8.1\,m Gemini South telescope and the GMOS
\citep[Gemini Multi-Object Spectrographs;][]{davies1997,gimeno2016} instrument.
Observations were conducted in the ``Poor Weather'' queue mode in April-June
2019, June-July 2021, and April-June 2022 (GS-2019A-Q-408, GS-2021A-Q-419, and
GS-2022A-Q-406). The exposure times ranged from 210 to 1800 seconds, with a
total of 45.84 hours on target.
The B600~l~mm$^{\rm{-1}}$ grating (G5323) and a 1$\farcs$5 slit were used with
a 2$\times$2 binning, resulting in a wavelength coverage in the range
[3200:5800]\,{\AA} at resolving power $R \sim 1,200$. The complete data
reduction was performed using the {\texttt{DRAGONS}}\footnote{\href{
    https://github.com/GeminiDRSoftware/DRAGONS}{https://github.com/GeminiDRSoftware/DRAGONS}.}
    software package \citep{dragons}.

\section{Stellar Parameters and Chemical Abundances}
\label{secatm}

The determinations of stellar atmospheric parameters (\teff, \logg, and \metal),
carbonicity (\cfe), and $\alpha$-to-iron ratios (\afe) for the stars observed as
part of spectroscopic follow-up were made using the n-SSPP
\citep{beers2014,beers2017}, a modified version of the SEGUE Stellar Parameter
Pipeline \citep[SSPP;][]{lee2008a,lee2008b,lee2011,lee2013}.  The code uses
photometric and spectroscopic information to calculate the atmospheric
parameters based on several different methods, including calibrations
with spectral line indices (from the \ion{Ca}{2} H and K lines for \metal),
photometric \teff\ predictions, and synthetic spectra matching. Further details
can be found in \citet{placco2018}.

The \cfe\, and \afe\, are estimated from the strength of the CH G-band molecular
feature at $\sim 4300$\,{\AA} and the \ion{Mg}{1} triplet at 5150--5200\,{\AA},
respectively. For the spectral fitting, the n-SSPP uses a grid of
synthetic spectra generated with the {\texttt{MARCS}} model atmospheres
\citep{gustafsson2008}, CH line lists from \citet{masseron2014}, and the
{\texttt{TURBOSPECTRUM}} code \citep{plez2012}, for the \cfe\ determination.
For \afe, a grid of synthetic spectra are created with the Kurcuz model
atmospheres\footnote{\href{http://kurucz.harvard.edu/grids.html}{http://kurucz.harvard.edu/grids.html}}
and line lists\footnote{\href{http://kurucz.harvard.edu/LINELISTS}{http://kurucz.harvard.edu/LINELISTS}}, which
is updated version from \citet{lee2011}.
The high-resolution synthetic
spectra have their resolution degraded to match that of the observed data.
The n-SSPP was able to estimate \teff, \logg, and \metal\ for all the 522 stars
observed as part of this work.  The \cfe\ and \afe\ abundance ratios were
estimated for 455\footnote{Most stars without carbon abundance determinations
were observed with CTIO/Blanco. There was an artifact at the exact same position
as the CH band head that affected some of the spectra and prevented reliable
spectral fits by the n-SSPP.} and 483 stars, respectively. 

The adopted stellar parameters and their uncertainties are calculated as the
biweight average of the individual accepted estimates and a robust estimate of
the scatter \citep[see details in][]{lee2008a}. For the \cfe\ and \afe\
abundance ratios, the values are determined by minimizing the distance between
the target and synthetic fluxes, using a reduced $\chi^2$ statistical criterion.
Uncertainties are estimated by the square root of diagonal elements of the
resulting covariance matrix obtained during the $\chi^2$ minimization
\citep{lee2013}. In addition, noise-injected synthetic spectra are used to
derive the uncertainty as a function of S/N \citep[see][for
details]{lee2011,lee2013}, by matching the S/N of the observed data.
The average uncertainties for the observed sample are 70~K for \teff\
($\sigma$\teff\ $\in$ \texttt{[40:175]}), 0.24~dex for \logg\ ($\sigma$\logg\
$\in$ \texttt{[0.1:0.4]}), 0.11~dex for \metal\ ($\sigma$\metal\ $\in$
\texttt{[0.05:0.20]}), and 0.21~dex for \cfe\, and \afe. 

\begin{figure}[!ht]
\epsscale{1.15}
\plotone{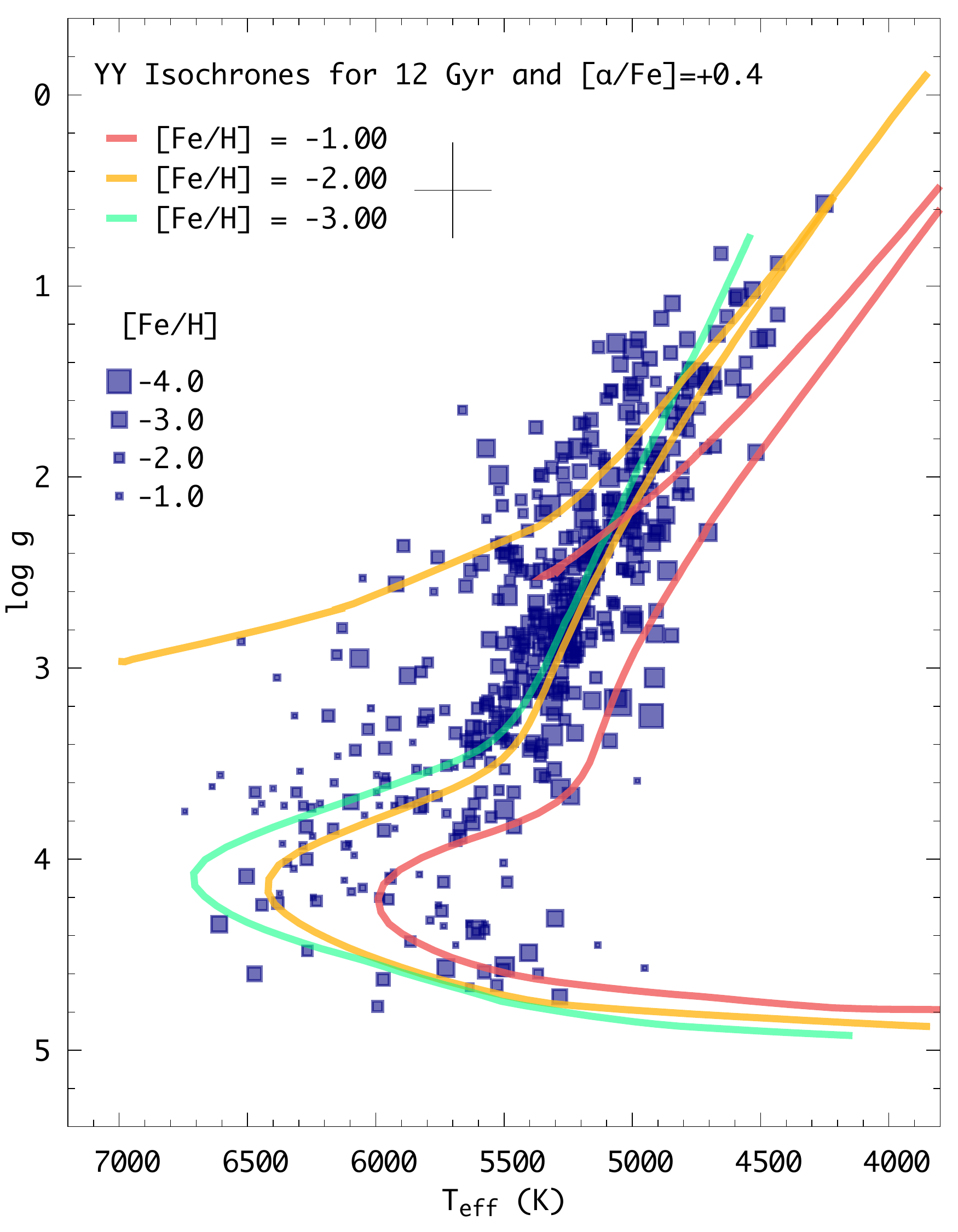}
\caption{Surface gravity vs. \teff\ diagram for the program stars, using the
parameters calculated by the n-SSPP, listed in Table~\ref{nsspp}. Typical
uncertainties are shown for reference. The point size is inversely proportional
to the metallicity. Also shown are the YY Isochrones \citep[12~Gyr, 0.8~M$_{\rm
\odot}$, \afe=$+$0.4;][]{demarque2004} for \metal\ = $-$1.0, $-$2.0, and $-$3.0,
and horizontal-branch tracks from \citet{dotter2008}.}
\label{isochrone}
\end{figure}

The adopted atmospheric parameters and abundance ratios for the sample are
listed in Table~\ref{nsspp}.
Also included in the table are the corrections for carbon abundances, based on
the stellar-evolution models presented in \citet{placco2014c}, the final \cfe,
and $A$(C)\footnote{$A$(C) = $\log(N_C/{}N_H) + 12$}, the latter two including
the corrections.  
Figure~\ref{isochrone} shows the \logg\ vs. \teff\ diagram for the sample,
compared with the YY isochrones for different metallicities \citep[12~Gyr,
0.8~M$_{\rm \odot}$, \afe=$+$0.4;][]{demarque2004}. The point sizes are
inversely proportional to the \metal\ values and typical uncertainties
for \logg\ ($\sim 0.25$~dex) and \teff\ ($\sim 150$~K) are also shown.
Based on the color selection from the \splus\ filters, it is expected that the
majority of the stars ($\sim78\%$) have temperatures in the
{\texttt{[4700:5700]}} K range. There is an overall agreement between
observations and the isochrones for \metal=$-2.0$ and $-3.0$, apart from a small
systematic offset of $\sim 50-100$~K for the spectroscopic temperatures. In
addition, it is evident that the majority of the higher metallicity stars
(smaller symbols, in particular \metal$\geq -1.0$, have \teff$\geq 5700$~K. This
will be further discussed in Section~\ref{improve}.

\begin{figure*}[!ht]
\epsscale{1.15}
\plotone{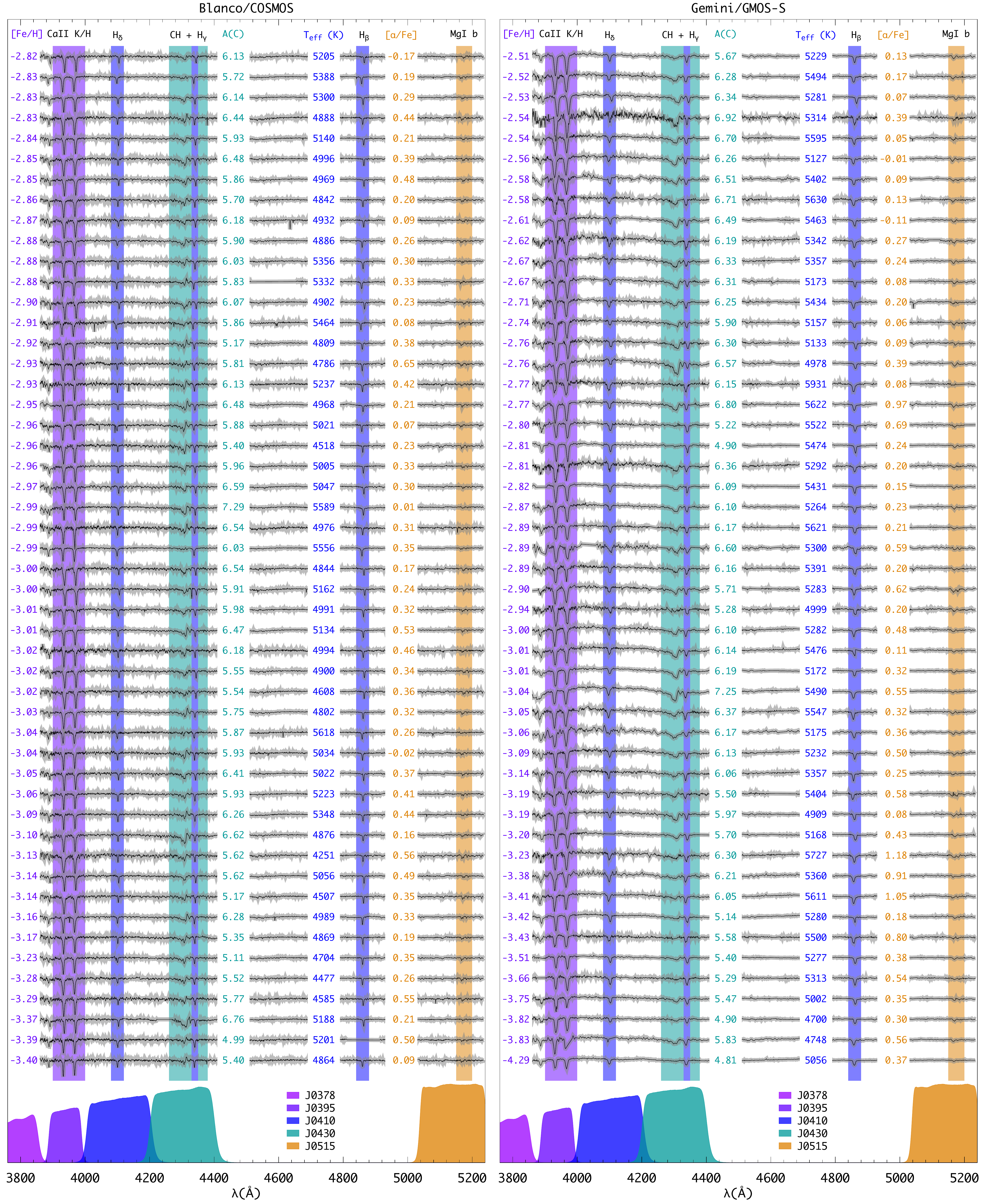}
\caption{Example spectra for 100 program stars observed with Blanco (left panel)
and Gemini-South (right panel), sorted by decreasing metallicity. The shaded
areas highlight absorption features of interest for the determination of the
stellar parameters and chemical abundances (see text for details). Also shown
are the values calculated by the n-SSPP, as well as the \protect\splus\ filters
that probe such features, with the exception of H$\gamma$ and H$\beta$.}
\label{spec}
\end{figure*}

\begin{figure*}[!ht]
\epsscale{1.15}
\plotone{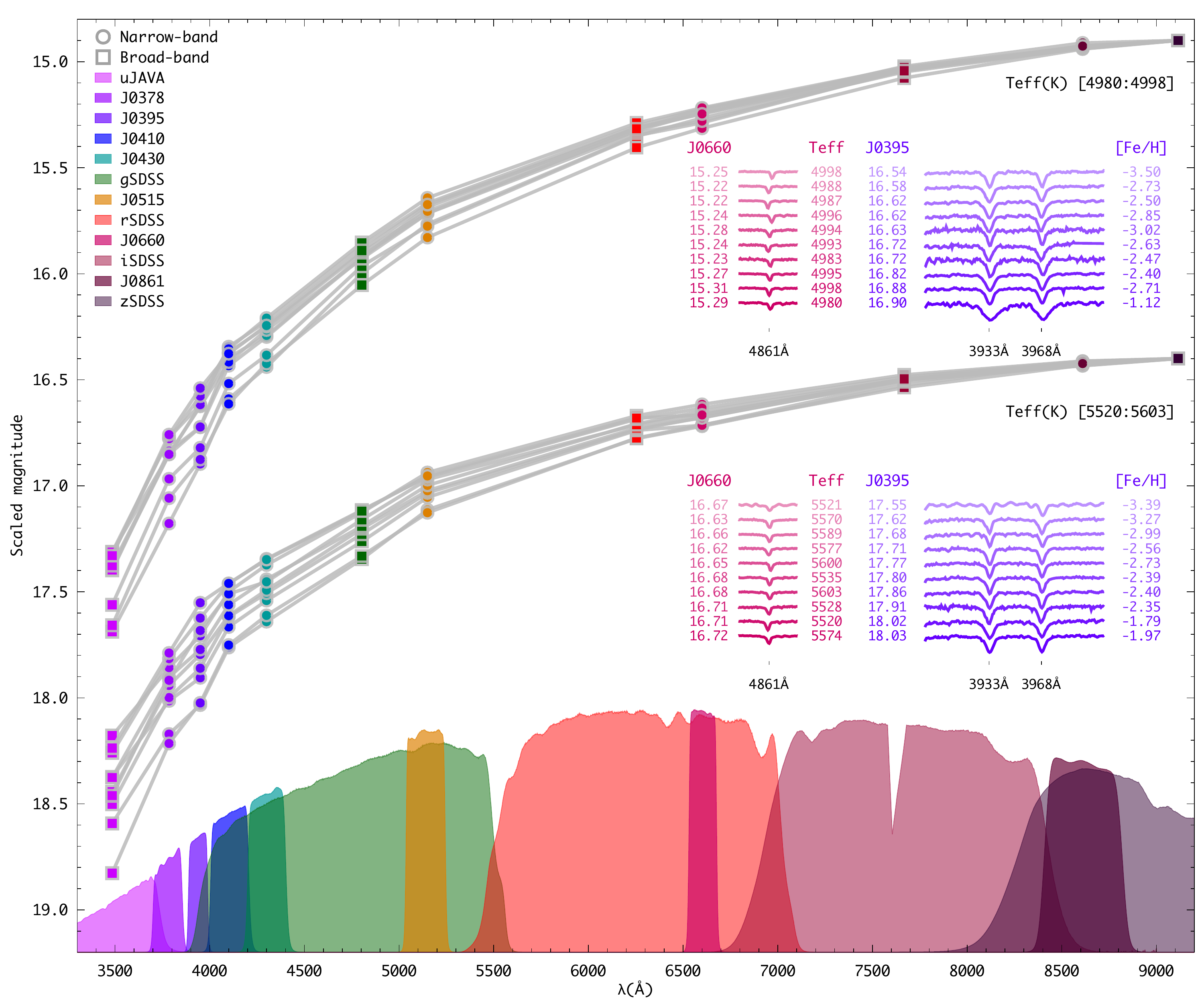}
\caption{Spectral energy distribution (SED) for 20 selected program stars from
Blanco and Gemini. Magnitudes were scaled to zSDSS=14.9 (top) and 16.4 (bottom).
The temperature range for each set is displayed right below the SEDs. The insets
show the observed spectra around the H$\beta$ and \ion{Ca}{2} HK absorption
features, sorted by their \jcaii\ scaled magnitude. Also shown are the \teff\
and \metal\ for each star and the \splus\ transmission curves. See text for
further details.}
\label{photospec}
\end{figure*}

Figure~\ref{spec} shows example spectra for the 100 most metal-poor stars
observed with Blanco (left panel) and Gemini-South (right panel), which have
both \cfe\ and \afe\ determined by the n-SSPP. Also shown are the adopted
parameters for each target (see Section~\ref{secatm} for details). The
absorption features of interest for the calculation of each parameter are
identified on the top of the panels. The shaded regions correspond to the
specific atmopspheric parameter or chemical abundance probed by the \splus\
filters outlined on the bottom of the panels. The spectra are sorted by
\metal. Despite the variation in \teff, it is possible to note the
overall decrease in the strength of the \ion{Ca}{2} features as the metallicity
decreases, as well as the increase in \teff, A(C), and \afe\ as their associated
absorption features strengthen.

\vspace{1.5cm}

\section{Analysis and Discussion}
\label{analysis}

\subsection{\metal\ sensitivity from narrow-band photometry}

As described in Section~\ref{secobs}, the \splus\ colors using the narrow-band
\jcaii\ filter are effective in separating different metallicity regimes (see
also Figure~\ref{binfull}). Hence, for a given temperature, the difference
between the \jcaii\ and \jhalf\ magnitudes should decrease as a function of
\metal\footnote{As an example, for two stars with the same \teff/\logg\ and at
the same distance, the one with the lowest metallicity would be brighter in
\jcaii.}. 
An attempt to illustrate and quantify this effect is shown in
Figure~\ref{photospec}. There are two sets of Spectral Energy Distributions
(SED) with ten stars each, in two narrow temperature intervals. The magnitudes
were scaled to an arbitrary zSDSS value in order to preserve their color indices
and allow for a star-to-star comparison of the sensitivity of the other
magnitudes to changes in stellar parameters.
The insets show sections of the observed spectra (sorted by their \jcaii\ scaled
magnitude) around the \ion{Ca}{2} HK and H$\beta$ features, as well as their
respective parameters (\metal\ and \teff) and scaled magnitudes (\jcaii\ and
\jhalf). Assuming that the three bluest filters carry most of the metallicity
information, it is possible to qualitatively see the larger variation in magnitudes, as
compared with the magnitudes from the redder filters. Under the assumption that
(most) of this variation is due to changes in metallicity, it is expected that
the sorted \jcaii\ magnitudes would naturally result in a \metal\ sequence.

\begin{figure}[!ht]
\epsscale{1.15}
\plotone{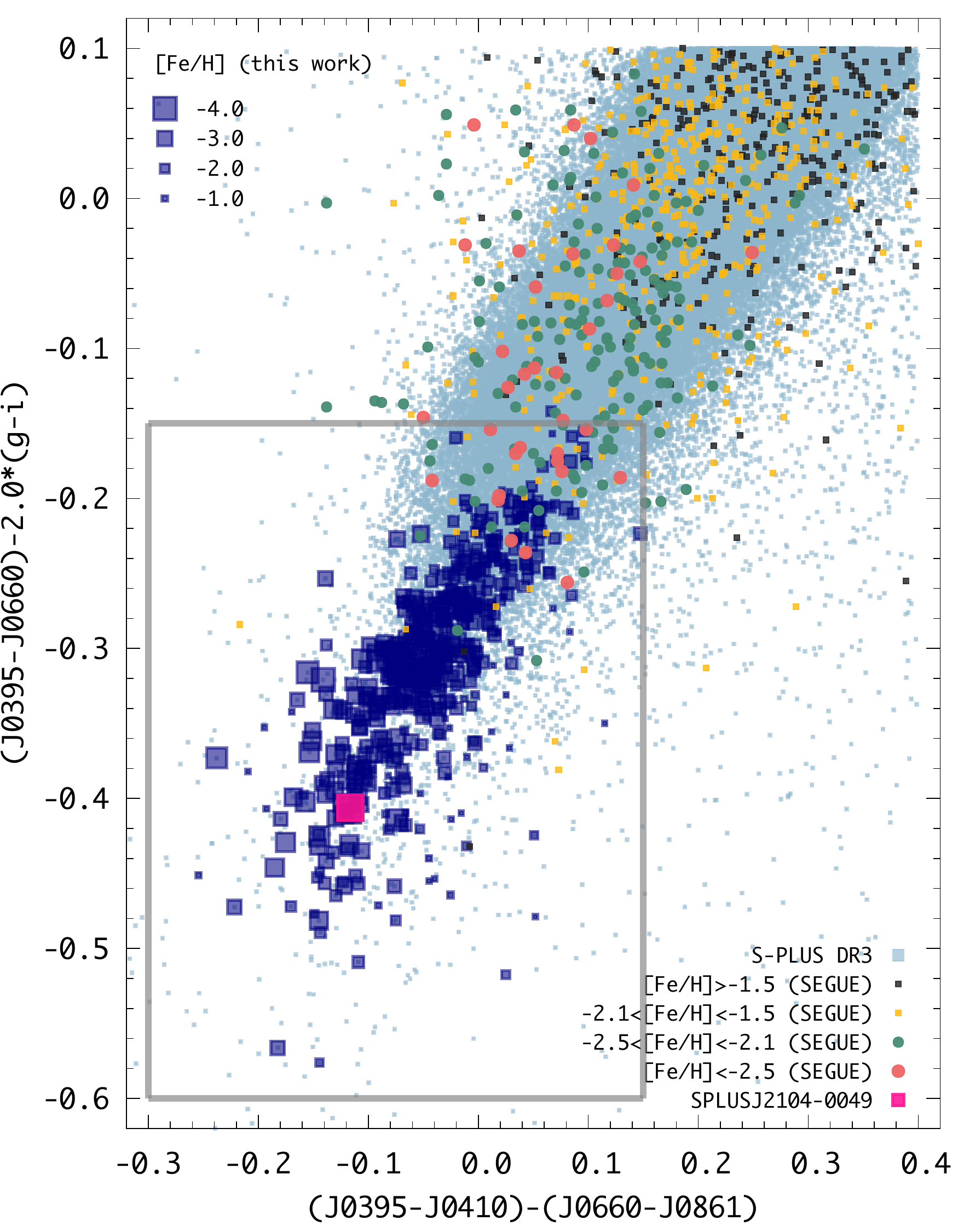}
\caption{\splus\ color-color diagram for \splus\ DR3 (light blue filled
squares), SDSS/SEGUE cross-match (divided in four metallicity intervals),
and the stars observed in this work (dark blue filled squares, with point sizes
inversely proportional to the metallicity). The gray rectangle outlines the
selection window for the spectroscopic follow-up, described in the text. Also
shown is \rave, an ultra metal-poor star identified in \splus\ by
\citet{placco2021b}.} 
\label{colorselect} \end{figure}

\begin{figure*}[!ht]
\epsscale{1.15}
\plotone{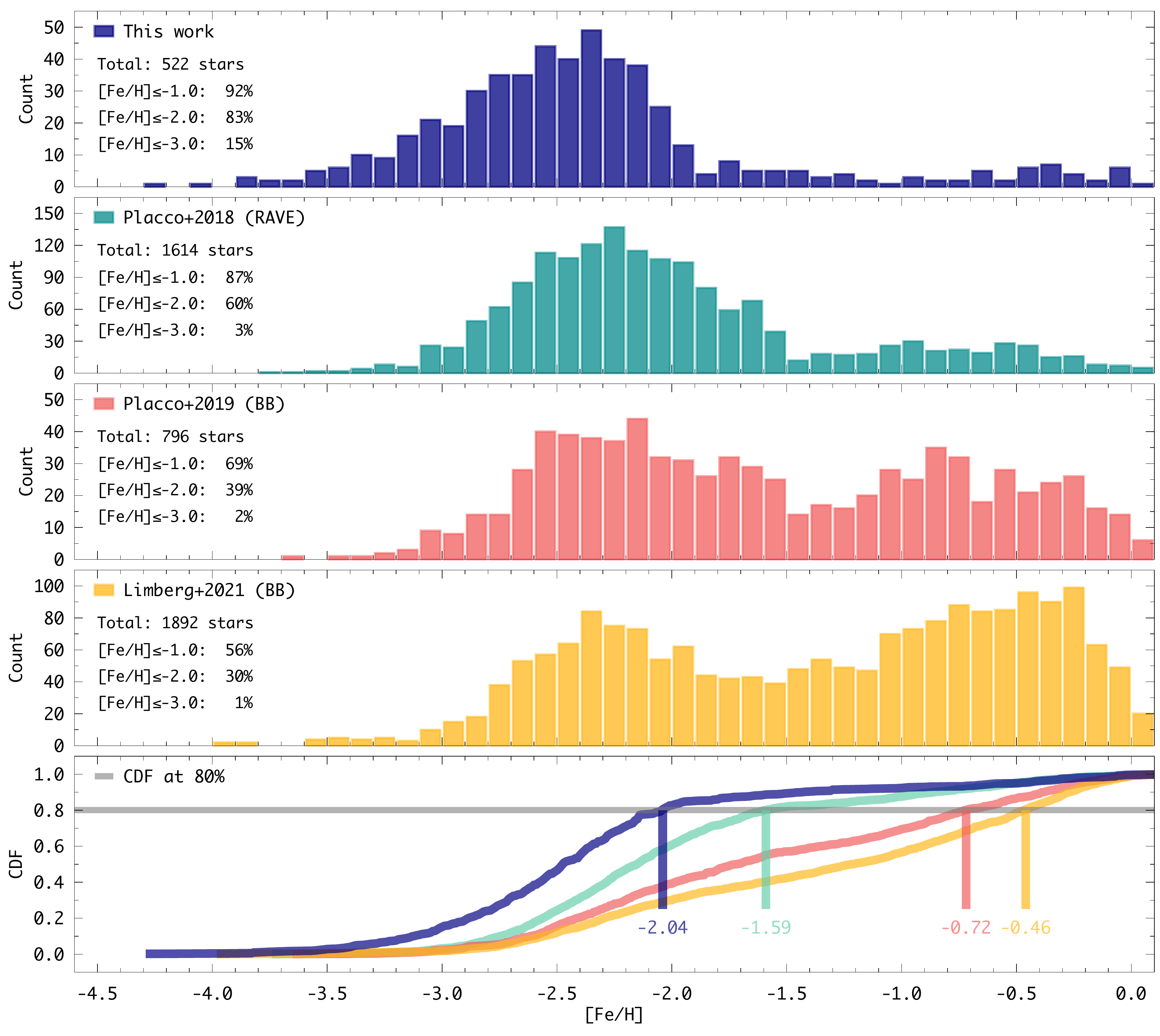}
\caption{Metallicity histogram for the program stars (top panel), compared with
the distributions from \citet{placco2018}, \citet{placco2019}, and
\citet{limberg2021} (middle panels). Each panel shows the total number of stars
and the fractions for different metalliticy regimes. The bottom panel shows the
cumulative distribution functions (CDF) for the three samples, marking the
\metal\ value for which they reach 80\%.}
\label{cdf}
\end{figure*}

\begin{figure*}[!ht]
\epsscale{1.15}
\plottwo{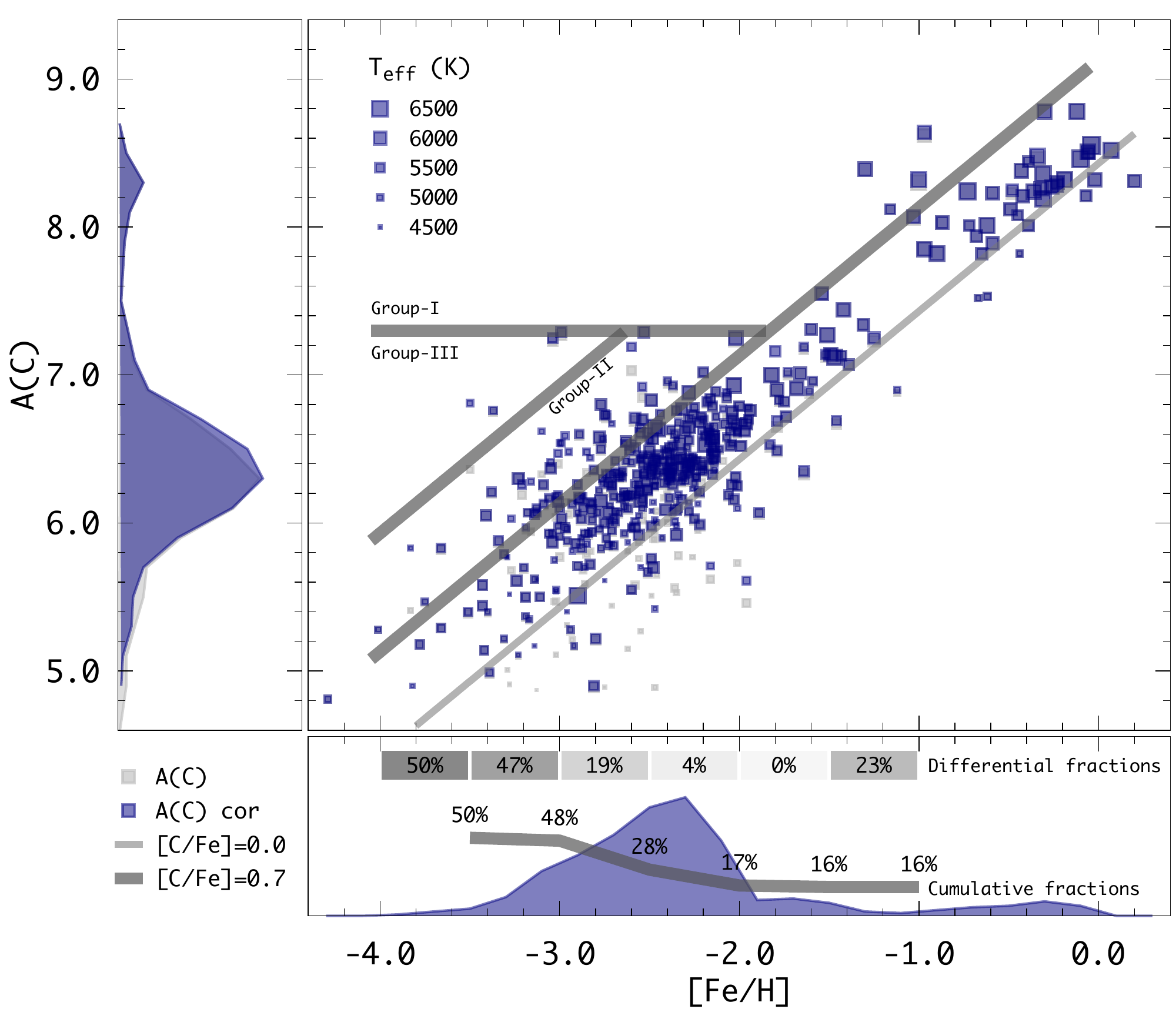}{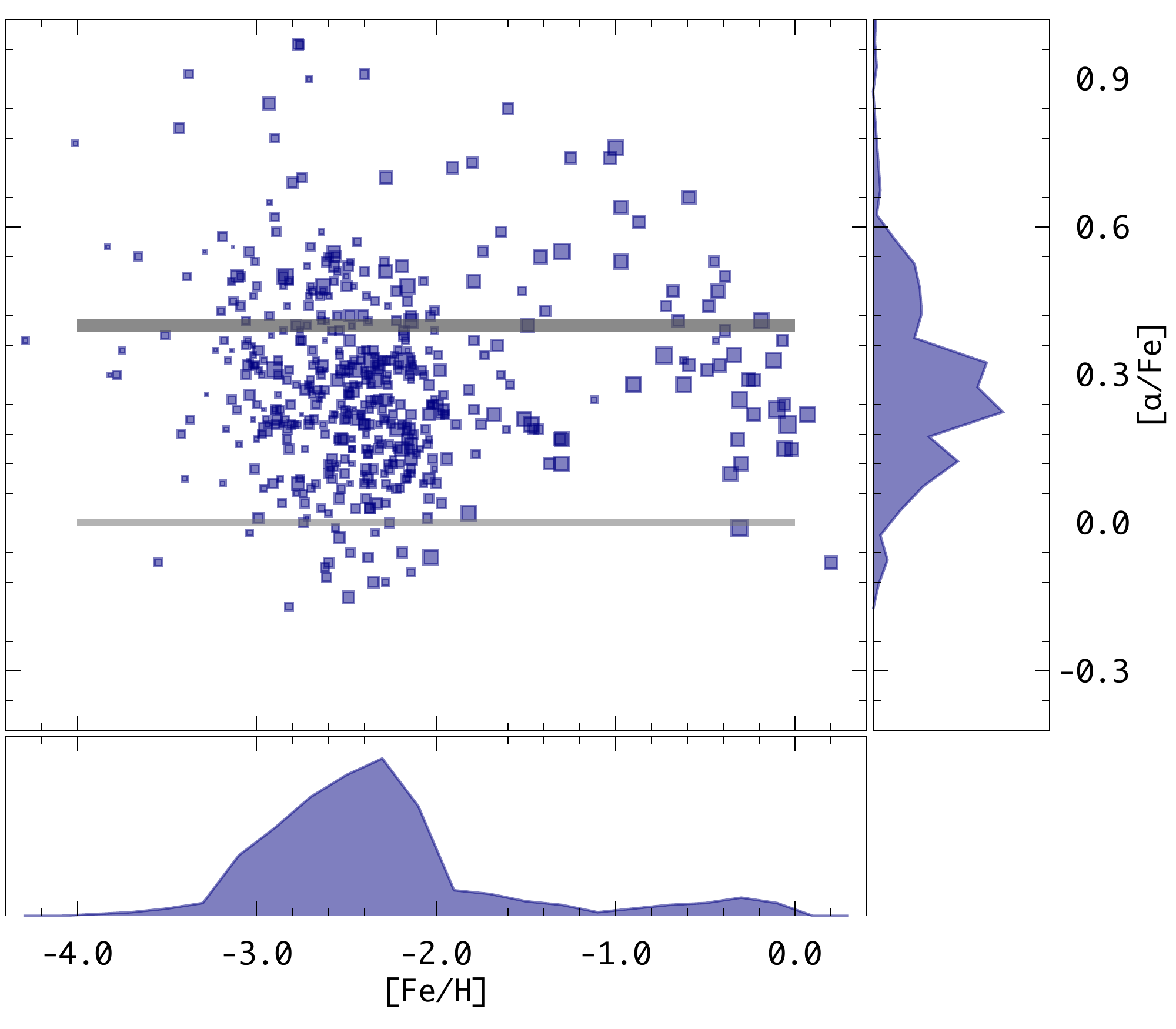}
\caption{Absolute carbon, ($A$(C), corrected as described in the text - left panel), and
$\alpha$-element abundance ratios, [$\alpha$/Fe] (right panel), as a function of the metallicity
calculated by the n-SSPP. The side and lower panels show the marginal
distributions. The solid line in the lower left panel shows the
cumulative CEMP fractions for the stars with $-3.5 \leq$\,\metal\,$\leq -1.0$
and the numbers of the top part of the panel are the differential fractions
for 0.5~dex \metal\ bins. Point sizes are proportional to \teff.}
\label{acfeh}
\end{figure*}

For the SEDs on the top set, all \teff\ values are within 18\,K, which
translates into a 0.09\,mag ($\sim 0.5\%$) variation in the \jhalf\ flux. As
expected, there is an increase of the \jcaii\ magnitude with \metal. The two
extremes of the \metal\ scale have a 0.36 magnitude difference in \jcaii, which
roughly translates into a 2.0~dex variation in \metal. At this temperature
range, the above variation in the \ion{Ca}{2} HK region flux is well above the
typical magnitude uncertainties from \splus, making it possible to have a good
metallicity discriminant.

To a certain degree, the same applies to the bottom set of SEDs, which has an
average \teff\ about 500\,K warmer than the top set, with a somewhat lager
dispersion of 83\,K. 
These warmer temperatures result in weaker absorption features when compared to
the cooler set for a given \metal, but are still larger than the typical
uncertainties in the measured flux.
For this group, the variation in the \jhalf\ magnitude is
very small (0.05\,mag), while the difference between the extremes in \jcaii\,
(0.48\,mag) still translates into a 1.5\,dex range in terms of \metal.

\subsection{Effectiveness of \splus\ color selection}

One of the main goals of this work is to improve the success rate of finding
low-metallicity stars from photometry, taking advantage of the metallicity
sensitivity of the \splus\ narrow-band filters. Figure~\ref{colorselect} shows
the color-color diagram for the \splus\ DR3 data, the cross-match between
\splus\ and SDSS/SEGUE (color-coded by metallicity range), and the 522 stars
observed in this work. The point size is inversely proportional to \metal.
The gray rectangle outlines the selection window for the spectroscopic
follow-up, as defined in Section~\ref{secobs}:
\colorx\,$\in$\,{\texttt{[$-$0.30:0.15]}} and
\colory\,$\in$\,{\texttt{[$-$0.60:$-$0.15]}}.
Also shown in the figure is \rave, the first ultra metal-poor star identified in
\splus, with \metal=$-4.03$\footnote{The n-SSPP estimated \metal=$-4.29$ from
the Gemini/GMOS medium-resolution spectrum.} \citep{placco2021b}.

There is, as expected, a strong correlation between metallicity and the position
of a star in this color-color diagram. However, that does not imply a direct
translation between these colors and \metal, as evidenced by the stars with
smaller points (higher metallicities) present towards negative colors. However,
these higher-metallicity stars can be filtered-out by another color combination
(see Section~\ref{improve} below for further details. What still holds true, as
set forth by Figure~\ref{binfull}, is the fact that the {\emph{fraction}} of
stars with \metal$\leq -2.0$ increases for decreasing \colory\, and \colorx.

A different procedure to assess the efficiency of the color selection is by
looking at the MDF of the observed stars and
compare it with previous attempts of following-up low-metallicity star
candidates.
The top panel of Figure~\ref{cdf} shows the MDF of the stars observed in this
work, compared with data from \citet{placco2018}, \citet{placco2019}, and
\citet{limberg2021}\footnote{Even though these three efforts have followed-up data
selected from different approaches, both had the goal of maximizing the number
of observed stars with \metal$\leq -2.0$. A comparison could also be made with
the work of \citet{aguado2019}, \citet{dacosta2019}, and \citet{galarza2022}.
However, these studies used photometric metallicities for their target
selection, as opposed to the color-color diagrams employed by this work.} in
the middle panels. The total number of stars and fractions for different
metallicity ranges are also shown for each sample. The bottom panel shows the
cumulative distribution function (CDF) for the three samples, indicating the
80th percentile value for \metal. It is possible to see that the \splus\ color
selection is far superior than the previous efforts in selecting
low-metallicity star candidates. The success rate for \metal$\leq -2.0$ is
$83^{+3}_{-3}\%$\footnote{Uncertainties in the fractions are represented by the
Wilson \\ score confidence intervals \citep{wilson1927}.}, as compared to 60\% in
\citet{placco2018}, 39\% in \citet{placco2019}, and 30\% in
\citet{limberg2021}. Finally, the fraction of stars with \metal$\leq -3.0$
(15\%) is higher than the fraction of stars with \metal$> -1.5$ (11\%), which
confirms the effectiveness of the \splus\ color window in selecting
low-metallicity stars.

\subsection{Carbon and $\alpha$-element abundances}

The carbon and $\alpha$-element abundances calculated by the n-SSPP can provide
further insight on the origin of the observed stars and serve as selection
criteria for high-resolution spectroscopic follow-up. 
Even though the carbon abundances in the SEGUE sample were not used for the
target selection in this work, the {\texttt{(g-i)}} color (and, to some extent,
also the \jcaii\ and \jhdel\ magnitudes) can be  affected by the presence of carbon
molecular bands in the spectrum. As a consequence of the decreased emerging flux
on the $g$ band, a fraction of cool carbon-enhanced stars may fall outside the
{\texttt{(g-i)}}\footnote{As an example, the star SDSS~J1327$+$3335,
with \teff=4530~K, \metal=$-3.38$, and \cfe=$+2.18$, has {\texttt{(g-i)}}=1.718
\citep{yoon2020}.} window defined in Section~\ref{secobs}. Regardless, the
abundances measured for the program stars allow for the calculation of the
fractions of Carbon-Enhanced Metal-Poor \citep[CEMP -
\cfe$\geq+0.7$;][]{aoki2007} stars as a function of the metallicity.

Figure~\ref{acfeh} shows the distribution of $A$(C) (left panel) and \afe\,
(right panel) for the sample stars as a function of the metallicity. The
auxiliary panels show the marginal distributions. The lower left panel shows
both the differential and cumulative CEMP fractions.  Solid lines mark the solar
values for the quantities, as well as the CEMP definition on the left panel and
the average \afe$= +0.4$ for the Galactic halo \citep{venn2004,kobayashi2006} in
the right panel. The point sizes are proportional to the temperature.  The CEMP
groups labeled are loosely defined based on the arguments presented in
\citet{yoon2016}, which were built upon the work of
\citet{spite2013,bonifacio2015,hansen2015}.
The average $\alpha$-element abundance for the sample (\afe$=+0.29$) is somewhat
lower than the typical value for stars with \metal$< -2.0$ and there is no
apparent trend with metallicities. 

Stars in the CEMP Group-I are believed to have acquired their carbon in a
binary system from a now-evolved companion that went through its AGB phase
\citep[extrinsic enrichment;][]{placco2014c}. Members of Groups-II and III, on
the other hand, carry in their atmosphere the carbon signature inherited from
its parent population \citep[intrinsic enrichment;][]{placco2014c}. The
distinction between Groups II and III lies on specific characteristics of the
massive stars that poluted the gas clouds from which the subsequent low-mass
stars were formed. For the sample presented in this work, there are 68 CEMP
stars, with $4$ stars in Group-I, $60$ in Group-II, and $4$ in Group-III.
Compared with the work of \citet{yoon2016}, the sample presented in this work
has an exceptionally low number of Group-I stars, which should be the majority
in the CEMP population. This may be partially\footnote{\citet{arentsen2022}
points out that, in general, low-metallicity star samples from medium-resolution
spectroscopy show a lower than expected fraction of CEMP Group-I stars when
compared to high-resolution samples.} a consequence of the {\texttt{(g-i)}}
restriction mentioned above and other \splus\ color selections, which would
exclude most of the higher metallicity and higher carbon abundance stars
associated with the Group-I.

The cumulative CEMP fractions showed in the lower panel of Figure~\ref{acfeh}
are consistently lower for \metal$< -1.0$ and $< -2.0$ when compared to other
empirical estimates in the literature, derived from samples with similar
spectral resolution \citep{placco2018,yoon2018,placco2019,limberg2021}. This may
be, once more, a consequence of the selection methods employed in this work. The
same applies to the differential fractions. The fractions are in better
agreement for the \metal$< -3.0$ regime, which could be due to the fact that
Group-III stars dominate the CEMP population in this metallicity regime
\citep[see][for a complete review of the CEMP fractions in the
literature]{arentsen2022}.

\subsection{Further improvements in the color selection}
\label{improve}

Even though the \splus\ color selections presented in this work are effective in
selecting low-metallicity stars, additional restrictions can be made in order to
decrease the number of stars with \metal$\geq -1.0$ for further spectroscopic
follow-up and targeting. Further inspection of Figure~\ref{isochrone} reveals
that 42\% of the stars with \teff$\geq 5900$~K have \metal$\geq -1.0$, while
only 3\% of the stars with \teff$< 5900$~K have \metal$\geq -1.0$. 

\begin{figure}[!ht]
\epsscale{1.15}
\plotone{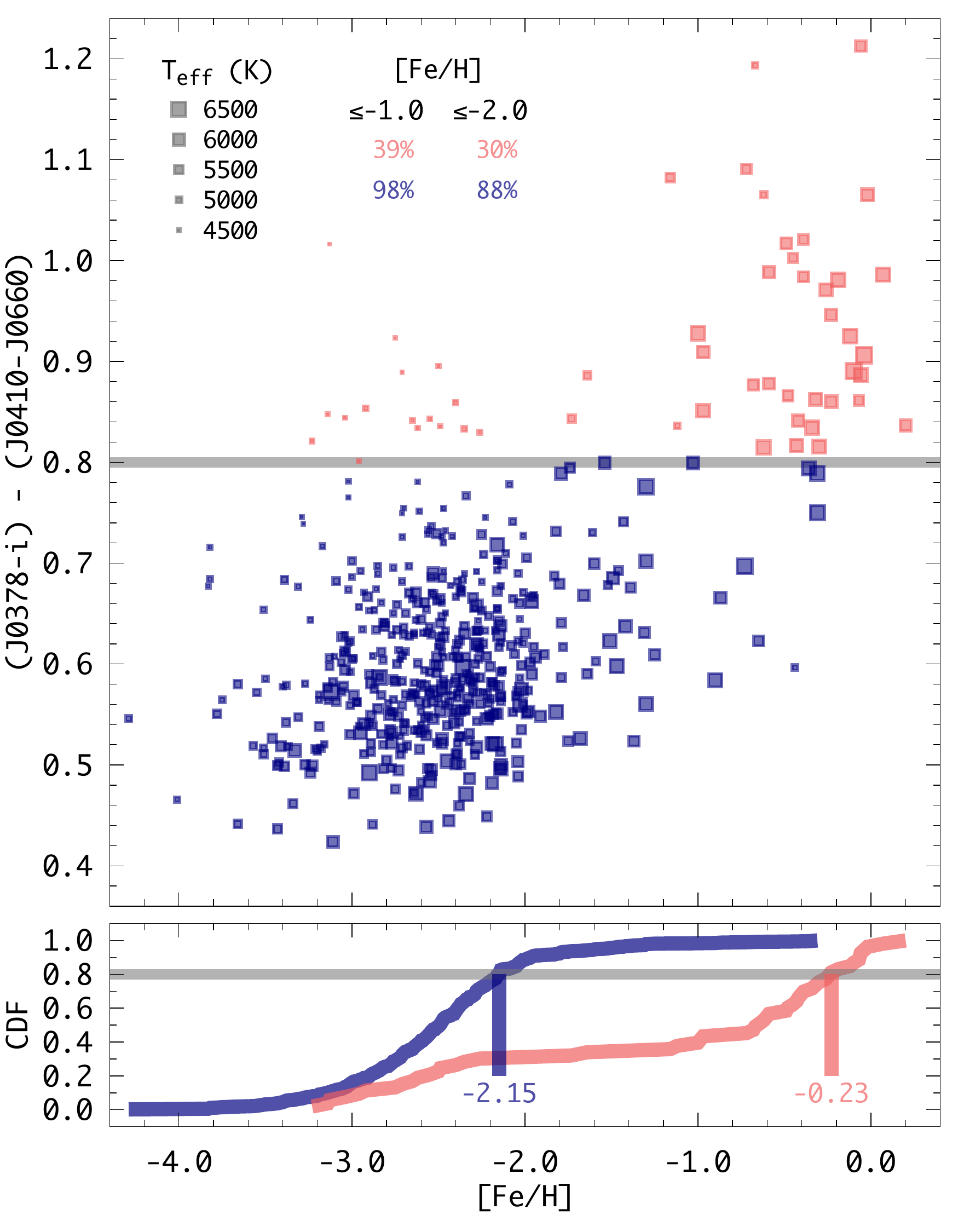}
\caption{Top panel: \colora\, as a function of the metallicity for the observed
sample, with the gray line marking the proposed color cut. Also shown are the
fractions of stars with \metal$\leq -1.0$ and $\leq -2.0$ for the two
subsamples. The point sizes are proportional to \teff. Bottom panel: cumulative
distribution functions (CDF) for the two subsamples, marking the \metal\ value
for which they reach 80\%.}
\label{cutfinal}
\end{figure}

An exploratory analysis was conducted to search for an additional \splus\, color
combination that would help remove the higher metallicity stars for the
continuation of the spectroscopic follow-up. One such candidate is \colora,
which contains the metallicity-sensitive {\texttt{J0378}} filter and the
temperature-sensitive {\texttt{(J0410-J0660)}} color index.  The top panel of
Figure~\ref{cutfinal} shows the behavior of this color as a function of the
metallicity. It is possible to see a very strong correlation between the two
quantities, and a tentative cut was made at \colora=0.80. The high color index
subsample (red points) is dominated by the higher metallicity stars, with only
30\% of the stars with \metal$\leq -1.0$ and 39\% with \metal$\leq -2.0$. For
the low color index subsample (blue points), the fractions are 98\% for stars
with \metal$\leq -1.0$ and 88\% with \metal$\leq -2.0$. The bottom panel shows
the CDFs for both subsamples, indicating the 80th percentile value for \metal.

This additional color restriction further improves the success rate for the
identification of stars with \metal$\leq -2.0$ from the \splus\, photometry. 
Even though there are low-metallicity stars with \colora$>0.8$, their fraction
is substantially smaller than for \colora$<0.8$. The stars at the high color
index subsample with \metal$\leq -2.0$ are all low-temperature (\teff$\lesssim
4800$~K) and low-carbon (\cfe$\leq 0.0$).

\section{Conclusions}
\label{final}

This work presented the medium-resolution spectroscopic follow-up of 522
low-metallicity star candidates selected from their \splus\ photometry.
By using metallicity-sensitive colors, the success rate found is
$92^{+2}_{-3}\%$ for \metal$\leq -1.0$, 
$83^{+3}_{-3}\%$ for \metal$\leq -2.0$, and  
$15^{+3}_{-3}\%$ for \metal$\leq -3.0$, including two ultra metal-poor stars
(\metal$\leq -4.0$). Based on the carbonicity determinations, there are 68 CEMP
stars in the sample, including $60$ Group-II and $4$ Group-III. Most of these
are already being followed-up with high-resolution spectroscopy in order to
determine their chemical abundance pattern and further understand their origin.
Based on the \metal\, determined in this work, a further color restriction is
proposed, which can potentially increase the fractions of stars with \metal$\leq
-1.0$ and $\leq -2.0$ to 98\% and 88\%, respectively.

The unpretentious color selection described in this work is not only extremely
effective in providing targets for further spectroscopic studies, but also
establishes a framework in which upcoming fiber-fed multiplex surveys can
benefit from in terms of targeting. These surveys will continue to provide the
individual pieces that constitute the cosmic puzzle that is the chemical
evolution of the universe.

\vspace{2.0cm}

The authors would like to thank the anonymous referee and members of the S-PLUS
community (Guilherme Limberg, Leandro Beraldo e Silva, and Simone Daflon) who
provided insightful comments on the manuscript.
The work of V.M.P. is supported by NOIRLab, which is managed by the Association
of Universities for Research in Astronomy (AURA) under a cooperative agreement
with the National Science Foundation.
F.A.-F. acknowledges funding for this work from FAPESP grants 2018/20977-2 and
2021/09468-1.
A.A. gratefully acknowledges funding from the European Research Council (ERC)
under the European Unions Horizon 2020 research and innovation programme (grant
agreement No. 834148).
Y.S.L. acknowledges support from the National Research Foundation (NRF) of
Korea grant funded by the Ministry of Science and ICT (NRF-2021R1A2C1008679).
The S-PLUS project, including the T80-South robotic telescope and the S-PLUS
scientific survey, was founded as a partnership between the Funda\c{c}\~{a}o de
Amparo \`{a} Pesquisa do Estado de S\~{a}o Paulo (FAPESP), the Observat\'{o}rio
Nacional (ON), the Federal University of Sergipe (UFS), and the Federal
University of Santa Catarina (UFSC), with important financial and practical
contributions from other collaborating institutes in Brazil, Chile (Universidad
de La Serena), and Spain (Centro de Estudios de F\'{\i}sica del Cosmos de
Arag\'{o}n, CEFCA). We further acknowledge financial support from the São Paulo
Research Foundation (FAPESP), the Brazilian National Research Council (CNPq),
the Coordination for the Improvement of Higher Education Personnel (CAPES), the
Carlos Chagas Filho Rio de Janeiro State Research Foundation (FAPERJ), and the
Brazilian Innovation Agency (FINEP).
The members of the S-PLUS collaboration are grateful for the contributions from
CTIO staff in helping in the construction, commissioning and maintenance of the
T80-South telescope and camera. We are also indebted to Rene Laporte, INPE, and
Keith Taylor for their important contributions to the project. From CEFCA, we
thank Antonio Mar\'{i}n-Franch for his invaluable contributions in the early
phases of the project, David Crist{\'o}bal-Hornillos and his team for their
help with the installation of the data reduction package \textsc{jype} version
0.9.9, C\'{e}sar \'{I}\~{n}iguez for providing 2D measurements of the filter
transmissions, and all other staff members for their support with various
aspects of the project.
Based on observations made at Cerro Tololo Inter-American Observatory at NSF’s
NOIRLab (NOIRLab Prop. IDs 2019B-0069, 2020A-0032, 2022A-210002; PI: V.
Placco), which is managed by the Association of Universities for Research in
Astronomy (AURA) under a cooperative agreement with the National Science
Foundation.
Based on observations obtained at the international Gemini Observatory (Program
IDs: GS-2019A-Q-408, GS-2021A-Q-419, GS-2022A-Q-406), a program of NSF’s
NOIRLab, which is managed by the Association of Universities for Research in
Astronomy (AURA) under a cooperative agreement with the National Science
Foundation. on behalf of the Gemini Observatory partnership: the National
Science Foundation (United States), National Research Council (Canada), Agencia
Nacional de Investigaci\'{o}n y Desarrollo (Chile), Ministerio de Ciencia,
Tecnolog\'{i}a e Innovaci\'{o}n (Argentina), Minist\'{e}rio da Ci\^{e}ncia,
Tecnologia, Inova\c{c}\~{o}es e Comunica\c{c}\~{o}es (Brazil), and Korea
Astronomy and Space Science Institute (Republic of Korea).
This research has made use of NASA's Astrophysics Data System Bibliographic
Services; the arXiv pre-print server operated by Cornell University; the
{\texttt{SIMBAD}} database hosted by the Strasbourg Astronomical Data Center;
and the online Q\&A platform {\texttt{stackoverflow}}
(\href{http://stackoverflow.com/}{http://stackoverflow.com/}).

\software{
{\texttt{awk}}\,\citep{awk}, 
{\texttt{dustmaps}}\,\citep{green2018}, 
{\texttt{DRAGONS}}\,\citep{dragons}, 
{\texttt{gnuplot}}\,\citep{gnuplot}, 
{\texttt{IRAF}}\,\citep{tody1986,tody1993}, 
{\texttt{matplotlib}}\,\citep{matplotlib}, 
{\texttt{n-SSPP}}\,\citep{beers2014}, 
{\texttt{numpy}}\,\citep{numpy}, 
{\texttt{pandas}}\,\citep{pandas}, 
{\texttt{sed}}\,\citep{sed},
{\texttt{stilts}}\,\citep{stilts}.
}

\facilities{
Blanco (COSMOS),
Gemini:South (GMOS)
}

\clearpage

\bibliographystyle{aasjournal}

\clearpage
\startlongtable

\begin{deluxetable*}{@{}ccrrrrclllr}
\tabletypesize{\scriptsize}
\tabletypesize{\tiny}
\tablewidth{0pc}
\tablecaption{Observing Details \label{telescope}}
\tablehead{
\colhead{Star Name}   &
\colhead{Star Name}   &
\colhead{$\alpha$}  &
\colhead{$\delta$}  &
\colhead{$l$}       &
\colhead{$b$}       &
\colhead{Date}        &
\colhead{Telescope}   &
\colhead{Instrument}  &
\colhead{Proposal ID} &
\colhead{Exp.}        \\
\colhead{(SPLUS)}     &
\colhead{(2MASS)}     &
\colhead{(J2000)}   &
\colhead{(J2000)}   &
\colhead{(deg)}     &
\colhead{(deg)}     &
\colhead{(UTC)}       &
\colhead{}            &
\colhead{}            &
\colhead{}            &
\colhead{(s)}         }
\startdata
J000445.50$+$010117.0 & J00044550$+$0101170 & 00:04:45.60 & $+$01:01:15.6 &  99.307 & $-$59.692 & 2020-12-26 &       Blanco & COSMOS &     2020A-0032 &  360 \\
J001736.44$+$000921.7 & J00173643$+$0009215 & 00:17:36.48 & $+$00:09:21.6 & 104.962 & $-$61.530 & 2020-12-27 &       Blanco & COSMOS &     2020A-0032 &  360 \\
J002554.41$-$305032.0 & J00255442$-$3050320 & 00:25:54.48 & $-$30:50:31.2 & 357.784 & $-$83.297 & 2020-12-27 &       Blanco & COSMOS &     2020A-0032 &  360 \\
J002712.10$-$313352.1 & J00271209$-$3133515 & 00:27:12.00 & $-$31:33:50.4 & 351.455 & $-$83.105 & 2020-12-25 &       Blanco & COSMOS &     2020A-0032 &  360 \\
J002712.43$+$010037.0 & J00271240$+$0100377 & 00:27:12.48 & $+$01:00:36.0 & 110.255 & $-$61.264 & 2020-12-26 &       Blanco & COSMOS &     2020A-0032 &  360 \\
J003555.86$-$420431.0 & J00355591$-$4204306 & 00:35:55.92 & $-$42:04:30.0 & 313.910 & $-$74.721 & 2021-01-11 &       Blanco & COSMOS &     2020A-0032 &  180 \\
J005037.10$-$315413.2 & J00503713$-$3154131 & 00:50:37.20 & $-$31:54:14.4 & 305.016 & $-$85.221 & 2020-12-28 &       Blanco & COSMOS &     2020A-0032 &  540 \\
J005037.17$-$340816.7 & J00503717$-$3408167 & 00:50:37.20 & $-$34:08:16.8 & 304.318 & $-$82.988 & 2021-01-11 &       Blanco & COSMOS &     2020A-0032 &   90 \\
J005208.98$-$004609.9 & J00520900$-$0046092 & 00:52:08.88 & $-$00:46:08.4 & 123.332 & $-$63.640 & 2019-10-17 &       Blanco & COSMOS &     2019B-0069 &  600 \\
J005428.84$-$300101.7 & J00542886$-$3001012 & 00:54:28.80 & $-$30:01:01.2 & 290.094 & $-$87.035 & 2020-12-25 &       Blanco & COSMOS &     2020A-0032 &  480 \\
\enddata
\tablecomments{(This table is available in its entirety in machine-readable form.)}
\end{deluxetable*}

\clearpage
\startlongtable

\begin{longrotatetable}
\begin{deluxetable*}{@{}c@{}c@{}c@{}c@{}c@{}c@{}c@{}c@{}c@{}c@{}c@{}c@{}c@{}c@{}c@{}c@{}c@{}c@{}c@{}c@{}c@{}c@{}c@{}c@{}c@{}}
\tabletypesize{\scriptsize}
\tabletypesize{\tiny}
\tablecaption{S-PLUS Photometry \label{tsplusphoto}}
\tablehead{
\colhead{Star Name}   &
\colhead{uJAVA}       & \colhead{$\sigma$}    &
\colhead{J0378}       & \colhead{$\sigma$}    &
\colhead{J0395}       & \colhead{$\sigma$}    &
\colhead{J0410}       & \colhead{$\sigma$}    &
\colhead{J0430}       & \colhead{$\sigma$}    &
\colhead{gSDSS}       & \colhead{$\sigma$}    &
\colhead{J0515}       & \colhead{$\sigma$}    &
\colhead{rSDSS}       & \colhead{$\sigma$}    &
\colhead{J0660}       & \colhead{$\sigma$}    &
\colhead{iSDSS}       & \colhead{$\sigma$}    &
\colhead{J0861}       & \colhead{$\sigma$}    &
\colhead{zSDSS}       & \colhead{$\sigma$}    
}
\startdata
J000445.50$+$010117.0 & 15.150 &  0.004 & 14.629 &  0.004 & 14.530 &  0.006 & 14.283 &  0.004 & 14.195 &  0.004 & 13.862 &  0.001 & 13.659 &  0.003 & 13.313 &  0.001 & 13.275 &  0.001 & 13.093 &  0.001 & 13.014 &  0.002 & 12.978 &  0.001 \\
J001736.44$+$000921.7 & 15.996 &  0.006 & 15.434 &  0.006 & 15.321 &  0.008 & 14.994 &  0.006 & 14.889 &  0.005 & 14.523 &  0.002 & 14.284 &  0.004 & 13.908 &  0.001 & 13.841 &  0.002 & 13.638 &  0.001 & 13.529 &  0.002 & 13.498 &  0.001 \\
J002554.41$-$305032.0 & 15.891 &  0.008 & 15.341 &  0.008 & 15.168 &  0.011 & 14.946 &  0.009 & 14.803 &  0.008 & 14.530 &  0.003 & 14.354 &  0.005 & 14.043 &  0.002 & 13.984 &  0.002 & 13.829 &  0.002 & 13.770 &  0.003 & 13.729 &  0.002 \\
J002712.10$-$313352.1 & 15.082 &  0.005 & 14.655 &  0.006 & 14.498 &  0.008 & 14.198 &  0.006 & 14.090 &  0.006 & 13.845 &  0.002 & 13.609 &  0.003 & 13.349 &  0.001 & 13.299 &  0.001 & 13.123 &  0.001 & 13.042 &  0.002 & 13.039 &  0.001 \\
J002712.43$+$010037.0 & 15.296 &  0.005 & 14.837 &  0.005 & 14.711 &  0.007 & 14.485 &  0.005 & 14.385 &  0.005 & 14.144 &  0.002 & 13.934 &  0.003 & 13.658 &  0.001 & 13.622 &  0.001 & 13.464 &  0.001 & 13.394 &  0.002 & 13.370 &  0.001 \\
J003555.86$-$420431.0 & 15.575 &  0.005 & 15.142 &  0.006 & 14.967 &  0.008 & 14.758 &  0.006 & 14.668 &  0.006 & 14.480 &  0.002 & 14.280 &  0.004 & 14.033 &  0.002 & 14.046 &  0.002 & 13.883 &  0.002 & 13.835 &  0.003 & 13.808 &  0.002 \\
J005037.10$-$315413.2 & 16.459 &  0.009 & 15.981 &  0.010 & 15.805 &  0.013 & 15.606 &  0.010 & 15.500 &  0.010 & 15.251 &  0.004 & 15.059 &  0.007 & 14.790 &  0.002 & 14.743 &  0.003 & 14.586 &  0.002 & 14.533 &  0.004 & 14.501 &  0.003 \\
J005037.17$-$340816.7 & 15.724 &  0.006 & 15.058 &  0.005 & 14.892 &  0.008 & 14.436 &  0.005 & 14.190 &  0.005 & 13.745 &  0.002 & 13.487 &  0.003 & 12.997 &  0.001 & 12.926 &  0.001 & 12.658 &  0.001 & 12.550 &  0.001 & 12.493 &  0.001 \\
J005208.98$-$004609.9 & 15.402 &  0.004 & 14.967 &  0.005 & 14.787 &  0.006 & 14.626 &  0.005 & 14.528 &  0.004 & 14.316 &  0.002 & 14.129 &  0.003 & 13.881 &  0.001 & 13.853 &  0.002 & 13.723 &  0.001 & 13.660 &  0.002 & 13.639 &  0.002 \\
J005428.84$-$300101.7 & 15.796 &  0.006 & 15.352 &  0.007 & 15.218 &  0.011 & 15.002 &  0.007 & 14.886 &  0.006 & 14.640 &  0.003 & 14.496 &  0.004 & 14.184 &  0.002 & 14.152 &  0.002 & 13.980 &  0.002 & 13.922 &  0.003 & 13.895 &  0.002 \\
\enddata
\tablecomments{(This table is available in its entirety in machine-readable form.)}
\end{deluxetable*}
\end{longrotatetable}

\clearpage
\startlongtable

\begin{deluxetable*}{@{}ccrrrrrrr}
\tabletypesize{\scriptsize}
\tabletypesize{\tiny}
\tablewidth{0pc}
\tablecaption{Stellar Parameters and Abundances from the n-SSPP \label{nsspp}}
\tablehead{
\colhead{Star Name}             &
\colhead{\teff}                 &
\colhead{\logg}                 &
\colhead{\metal}                &
\colhead{\cfe}                  &
\colhead{$\Delta$\cfe\tablenotemark{a}}       &
\colhead{\cfe$_{\rm cor}$\tablenotemark{b}}   &
\colhead{$A$(C)$_{\rm cor}$\tablenotemark{c}} &
\colhead{\afe}                  \\
\colhead{(SPLUS)}               &
\colhead{(K)}                   &
\colhead{(cgs)}                 &
\colhead{}                      &
\colhead{}                      &
\colhead{}                      &
\colhead{}                      }
\startdata
J000445.50$+$010117.0 & 5227 & 2.56 & $-$2.37 & $+$0.85 & $+$0.02 & $+$0.87 & $+$6.93 & $+$0.30 \\
J001736.44$+$000921.7 & 4993 & 2.19 & $-$2.63 & $+$0.61 & $+$0.01 & $+$0.62 & $+$6.42 & $+$0.20 \\
J002554.41$-$305032.0 & 5186 & 1.72 & $-$2.21 & $+$0.05 & $+$0.35 & $+$0.40 & $+$6.62 & $+$0.31 \\
J002712.10$-$313352.1 & 5257 & 2.74 & $-$2.27 & $+$0.24 & $+$0.01 & $+$0.25 & $+$6.41 & $+$0.12 \\
J002712.43$+$010037.0 & 5394 & 3.41 & $-$2.29 & $+$0.21 &    0.00 & $+$0.21 & $+$6.35 & $+$0.53 \\
J003555.86$-$420431.0 & 5645 & 3.38 & $-$2.53 & $+$0.51 &    0.00 & $+$0.51 & $+$6.41 & $+$0.31 \\
J005037.10$-$315413.2 & 5384 & 3.06 & $-$2.32 & $+$0.26 & $+$0.01 & $+$0.27 & $+$6.38 & $+$0.15 \\
J005037.17$-$340816.7 & 4434 & 0.88 & $-$2.71 & \nodata & \nodata & \nodata & \nodata & $+$0.47 \\
J005208.98$-$004609.9 & 5556 & 2.85 & $-$2.99 & $+$0.58 & $+$0.01 & $+$0.59 & $+$6.03 & $+$0.35 \\
J005428.84$-$300101.7 & 5497 & 3.53 & $-$2.05 & $+$0.14 &    0.00 & $+$0.14 & $+$6.52 & $+$0.01 \\
\enddata
\tablenotetext{a}{Carbon correction from \citet{placco2014c}.}
\tablenotetext{b}{Corrected carbon-to-iron ratio.}
\tablenotetext{c}{Corrected absolute carbon abundance.}
\tablecomments{(This table is available in its entirety in machine-readable form.)}
\end{deluxetable*}

\end{document}